\documentclass[prd,nofootinbib,notitlepage,onecolumn,10pt]{revtex4-1}

\pdfoutput=1

\usepackage{amssymb}
\usepackage{amsmath}
\usepackage{graphicx,subfigure}
\usepackage{longtable}
\usepackage{verbatim}
\usepackage{amsfonts}
\usepackage{color}
\usepackage{float}
\usepackage{booktabs,multirow,array}
\usepackage[table]{xcolor}
\usepackage{booktabs}
\usepackage{microtype}

\usepackage{graphicx,amsmath,amssymb,epsfig,verbatim,mathrsfs,array,layout,textcomp,latexsym,xspace}
\usepackage{multirow}

\begin{document}

\begin{flushright}
DO-TH 16/06 \\
SI-HEP-2016-11
\end{flushright}

\title{Zooming in on $B\to K^\ast \ell\ell$ decays at low recoil}

\author{Simon Bra\ss{}}
\affiliation{Theoretische Physik 1,
Naturwissenschaftlich-Technische Fakult\"at,
Universit\"at Siegen, D-57068 Siegen, Germany}

\author{Gudrun Hiller}
\affiliation{Institut f\"ur Physik, Technische Universit\"at Dortmund, D-44221 Dortmund, Germany}
\author{Ivan Ni\v sand\v zi\'c}
\affiliation{Institut f\"ur Physik, Technische Universit\"at Dortmund, D-44221 Dortmund, Germany}
\begin{abstract}
We analyse $B\to K^\ast \ell\ell$ decays in the region of low hadronic recoil, where an operator product expansion (OPE) in $1/m_b$ applies.
Using a local model for charm  contributions  based on $e^+ e^- \to hadrons$ against the OPE provides a data-driven method to access the limitations to the OPE's accuracy related to binnings in the dilepton mass.
Model-independent fits to  $B\to K^\ast \mu \mu $ low recoil angular observables exhibit presently only small sensitivity to different charm models. They give similar results as the fits based on the OPE, and are in agreement with the standard model, but leave also room for new physics. Measurements with resolution small enough to probe charm resonances would be desirable.
\end{abstract}

\maketitle

\section{Introduction}

Rare (semi-)leptonic decays induced by $b\to s\ell\ell$ flavor-changing neutral current (FCNC) transitions are highly suppressed in the Standard Model (SM)
and therefore sensitive to effects from non-standard interactions. The corresponding exclusive $B$-meson decays  have been investigated
by the experimental collaborations LHCb~\cite{Aaij:2013iag, Aaij:2015oid}, CMS~\cite{Khachatryan:2015isa},  CDF~\cite{Aaltonen:2011ja}, Belle~\cite{Wei:2009zv} and BaBar~\cite{Aubert:2008bi}.
Recently, LHCb presented updated results on the full angular distribution of the process $B\to (K^{\ast}\to K\pi)\mu\mu$ from the data sample that corresponds to the total integrated luminosity of $3\,\text{fb}^{-1}$~\cite{Aaij:2015oid}. 
Further significant improvements in the precision of the measurements are expected in the ongoing LHC  run and the  LHCb upgrade~\cite{Bediaga:2012py}, as well as future machines~\cite{Aushev:2010bq}.

To fully exploit the forthcoming measurements it requires sufficient understanding of the long-distance backgrounds within the SM and/or the methods to disentangle them from the short-distance effects that might carry information about beyond-the-standard-model (BSM) physics. The non-perturbative QCD dynamics in the matrix elements of the local quark currents between the initial and final meson states is parameterised by hadronic transition form factors.
The latter can be computed in the region of
low hadronic recoil, which is the focus of this work, in the framework of Lattice QCD. Recent progress for $B \to K^*$ and $B_s \to \phi$ transitions
has been reported in~\cite{Horgan:2013hoa}. 

Another  important irreducible class of  long-distance phenomena stems from the resonances that are induced by four-quark operators.
The model-independent description of these effects, based on first theory principles, is currently not available and one needs to rely on models, which ideally can then be tested, {\it  i.e.}\@ compared to data.
One such tool is the low recoil Operator Product Expansion (OPE), in which the  effects of the non-local matrix elements of the four-quark operators can be computed in terms of local matrix elements in powers of $1/Q$~\cite{Grinstein:2004vb, Beylich:2011aq}. 
Here, the hard scale is provided by $Q \sim (\smash{\sqrt{q^2}},m_b)$, where $\smash{\sqrt{q^2}}$ denotes the invariant mass  of the dileptons which at low recoil is of the order of the $b$-quark mass, $m_b$~\cite{Buchalla:1998mt}.

The QCD equations of motion can be used to derive the improved Isgur-Wise relations~\cite{Grinstein:2002cz}  between form factors, valid at leading order in $1/m_b$~\cite{Bobeth:2010wg}. Together with the OPE, these relations imply  universality, that is,  independence on the polarization of the final state hadron,  of the transversity amplitudes in the high-$q^2$ region~\cite{Bobeth:2010wg}. This feature enables the construction of  observables free of  short-distance dependence assuming no significant  right-handed currents~\cite{Bobeth:2010wg, Bobeth:2012vn}. One can then use these observables to extract ratios of form factors independently of the underlying short-distance physics
to be used directly  in SM tests~\cite{Hambrock:2012dg, Hambrock:2013zya}.
Uncertainties due to  next-to-leading order $1/m_b$-corrections to the universality relations turn out to be parametrically suppressed,
 at percent level~\cite{Bobeth:2010wg, Bobeth:2012vn}.

In the region above the $\bar{c}c$-threshold, the charm loop effects turn into the nonperturbative resonant spectrum in 
$B\to K^{(\ast)}\ell\ell$ distributions, that shows up as peaks from narrow resonances $J/\psi, \psi(2S)$ and "wiggles" for higher $1^{--}$ states above the $\bar{D}D$-threshold \cite{Kruger:1996cv,Ali:1999mm}. 
 While the narrow  resonances are removed by kinematic cuts and are not directly relevant at low recoil anyway,  the wiggles, observed  in $B^+\to K^+\mu\mu$ decays~\cite{Aaij:2013pta}, constitute a background  not captured locally by the OPE.
 Since the local resonance structure is a non-perturbative effect, it is not revealed at any order of the perturbative OPE. 
 The amount of duality violation was  investigated in a toy model in Ref.~\cite{Beylich:2011aq}; one expects that the OPE gives  a reasonably good description  for binned observables. 
In view of the increasing precision it is therefore important to understand this quantitatively  for given bin position and size.
   
To assess the performance of the OPE we 
exploit  existing data on $B\to (K^{\ast}\to K\pi)\mu\mu$ angular distributions~\cite{Aaij:2015oid} in different binnings 
\begin{equation} \label{eq:bins}
[15 - 19] \, \mbox{GeV}^2 \, , \quad [15 - 17], [17 - 19] \, \mbox{GeV}^2 \, , \quad [15 - 16], \dots ,  [18 - 19] \, \mbox{GeV}^2 \, ,
\end{equation}
allowing to zoom in with resolution $\Delta q^2=4, 2$ and $1\, \mbox{GeV}^2$, respectively.
The differential branching fraction is available in the two larger binnings only~\cite{LHCb-BR}.
As we assume new physics at the electroweak scale or higher,
{\it a binning-related effect is due to resonances, not BSM physics}.

The plan of the paper is as follows: 
In Section~\ref{sec:general} we give the effective Hamiltonian and $B \to K^*(\to K \pi) \mu\mu$ angular observables.
In Section~\ref{sec:hiq2} we briefly review the low recoil OPE and the Kr\"uger-Sehgal approach~\cite{Kruger:1996cv} modelling resonance distributions locally and to be used as a test-case against the OPE.
Section~\ref{sec:wiggles} is devoted to the details of  such tests and gives results of a global fit for resonance parameters.
In Section~\ref{sec:SM} we present the outcome of the  global fit for  the BSM Wilson coefficients and provide estimates of OPE uncertainties.
We conclude in Section~\ref{sec:con}. Auxiliary information can be seen in three appendices.

\section{$B \to K^* \ell \ell$ generalities}
 \label{sec:general}

We briefly review the effective Hamiltonian in Section~\ref{The effective Hamiltonian} and the basics of the $B\to K^* (\to K \pi) \ell\ell$ angular observables in Section~\ref{The definitions of the angular distributions}, respectively. 
\subsection{The effective Hamiltonian}\label{The effective Hamiltonian}
We employ in this work the effective weak Hamiltonian description for  $b\to s\ell\ell$ transitions
\begin{equation}
\mathcal{H}_{\text{eff}}=-\frac{4G_F}{\sqrt{2}}  V_{tb}^{\vphantom{*}} V_{ts}^*
\sum_i\mathcal{C}_i(\mu)\mathcal{O}_i(\mu)+h.c. \, ,
\end{equation}
where $\mathcal{O}_i$ and $\mathcal{C}_i$  denote the dimension-six  operators and their Wilson coefficients, respectively. $\mu$ is an (arbitrary) renormalization scale and
$V_{ij} $ are CKM matrix elements.
We use the basis of the four-quark operators $\mathcal{O}_{1,\ldots 6}$ introduced in Ref.~\cite{Chetyrkin:1996vx}, {\it  i.e.}\@ the so called CCM-basis
\begin{equation}
\label{EffHamiltonian1}
\begin{split}
\mathcal{O}_1&=\vphantom{\sum_q}(\bar{s}_L\gamma_\mu T^a c_L)( \bar{c}_L \gamma^\mu T^a b_L),\quad
\mathcal{O}_2=(\bar{s}_L\gamma_\mu c_L) (\bar{c}_L \gamma^\mu b_L), \\
\mathcal{O}_3&=(\bar{s}_L\gamma_\mu b_L)\sum_q(\bar{q}\gamma^\mu q),\quad
\mathcal{O}_4=(\bar{s}_L\gamma_\mu T^a b_L)\sum_q(\bar{q}\gamma^\mu T^a q),\\
\mathcal{O}_5&=(\bar{s}_L\gamma_\mu \gamma_\nu \gamma_\rho b_L)\sum_q(\bar{q}\gamma^\mu \gamma^\nu \gamma^\rho q),\quad
\mathcal{O}_6=(\bar{s}_L\gamma_\mu \gamma_\nu \gamma_\rho T^a b_L)\sum_q(\bar{q}\gamma^\mu \gamma^\nu \gamma^\rho T^a q).
\end{split}
\end{equation}
Here, $T^a$ denote the generators of QCD and the sums are over active quark flavors $q=u,d,s,c,b$.

The photon (gluon) penguin operators $\mathcal{O}_7 (\mathcal{O}_8)$ and the semileptonic operators $\mathcal{O}_{9,10}$ are given as
\begin{equation}\label{EffHamiltonian2}
\begin{split}
&\mathcal{O}_7=\frac{e}{16\pi^2}m_b(\bar{s}\sigma^{\mu\nu}P_R b)F_{\mu\nu},\quad \mathcal{O}_8=\frac{g_s}{16\pi^2}m_b(\bar{s}\sigma^{\mu\nu}P_R  T^ab)G^a_{\mu\nu},\\
& \mathcal{O}_9=\frac{e^2}{16\pi^2}(\bar{s}\gamma^{\mu}P_L b)(\bar\ell\gamma_\mu\ell),\quad
\mathcal{O}_{10}=\frac{e^2}{16\pi^2}(\bar{s}\gamma^{\mu}P_L b)(\bar \ell\gamma_\mu\gamma_5\ell) \, ,
\end{split} 
\end{equation}
with chiral projectors $\smash{P_{L,R}=(1\mp \gamma_5)/2}$. The mass of the b-quark is the running mass in the $\overline{MS}$ scheme at  the scale $\mu$. We  neglect  the mass of the $s$-quark as well as the ones of the leptons and CKM-subleading contributions proportional to $V_{ub}^{\vphantom{*}} V_{us\vphantom{b}}^*$.

\subsection{The angular distribution}\label{The definitions of the angular distributions}

The full  angular distribution \cite{Kruger:1999xa} of $B\to K^* (\to K \pi) \ell\ell$  decays   \footnote{Since we are  concerned with CP-averaged quantities only we do not distinguish in the notation between mesons and their CP-conjugates.
}
 can be written as
\begin{equation}
\begin{split}
\frac{d^4\Gamma}{dq^2d\cos\theta_\ell d\cos\theta_K d\phi}&=\frac{3}{8\pi}J(q^2,\cos\theta_\ell,\cos\theta_K,\phi)  \, ,
 \label{Angular distributions}
\end{split}
\end{equation}
where
\begin{equation}
\begin{split}
J(q^2,\theta_\ell,\theta_K,\phi)&=J_1^s \sin^2 \theta_K + J_1^c\cos^2 \theta_K + (J_2^s\sin^2\theta_K+ J_2^c\cos^2\theta_K)\cos 2\theta_\ell\\
&+ J_3\sin^2\theta_\ell\sin^2\theta_K\cos 2\phi + J_4 \sin2\theta_\ell \sin 2\theta_K\cos\phi\\
&+ J_5 \sin\theta_\ell \sin 2\theta_K\cos\phi + J_6\cos\theta_\ell\sin^2\theta_K\\
&+ J_7\sin\theta_\ell\sin 2\theta_K\sin\phi + J_8\sin 2\theta_\ell \sin 2\theta_K \sin\phi\\
&+ J_9 \sin^2\theta_\ell\sin^2\theta_K\sin 2\phi \, . 
\end{split}
\end{equation}
We adopt the definitions of the angles  from~\cite{Bobeth:2012vn}, that is, 
 $\theta_\ell$ is the angle between the $\mu^-$ and the $B$  in the rest frame of the muon pair,
$\theta_K$ is the angle between the kaon and the negative direction of flight of the $B$ in the $K\pi$-rest  frame and $\phi$ is the angle between the normals to the planes spanned by the $K\pi$ and $\mu^+\mu^-$pairs in the rest frame of the $B$.
The angular coefficients $J_i=J_i(q^2)$  can be expressed in terms of  transversity amplitudes $\mathcal{A}_i^{a}$, {\it i.e.,}\@ the transition amplitudes with specified polarization of the final vector meson, $i=\perp,0,\parallel$ and the lepton pair, $a=L, R$, see Appendix A. 
Neglecting the mass of the leptons  the number of independent angular coefficients is eight \cite{Egede:2010zc}.
The angular distribution  for the CP-conjugate decay, $d^4 \bar \Gamma$, can be obtained by replacing in $J$ all angular coefficients 
$J_{1,2,3,4,7} \to + \bar J_{1,2,3,4,7} $ and $J_{5,6,8,9} \to - \bar J_{5,6,8,9}$, where $\bar{J}_i$ equal $J_i$ with  the weak phases complex-conjugated~\cite{Bobeth:2008ij}.

We consider the  observables $F_L$, the fraction of the longitudinally polarized $K^\ast$ mesons, and the CP-averaged ratios
\begin{equation}
S_i \equiv \frac{J_i+\bar{J}_i}{d\Gamma/dq^2+d\bar{\Gamma}/dq^2}  \,  .\label{S_i}
\end{equation}
The forward-backward asymmetry in the lepton angles can be identified  as $A_{\rm FB} =  S_{6}$.
Due to the different definitions of angles and normalization of the $J_i^j$  the following relations to the conventions used by  LHCb~\cite{Aaij:2015oid, Gratrex:2015hna} hold
\begin{equation}
F_L=F_L^{\text{LHCb}} \, , \quad  S_{3,5,7,9}=\frac{3}{4}S_{3,5,7,9}^{\text{LHCb}} \, ,  \quad S_{4,8}=-\frac{3}{4} S_{4,8}^{\text{LHCb}}  \, , \quad A_{\rm FB}=-A_{\rm FB}^{\text{LHCb}}.\label{translation between conventions}
\end{equation} 

Furthermore, endpoint relations apply, which are based on general grounds \cite{Hiller:2013cza} and hold irrespective of the underlying electroweak model
\begin{align} \label{eq:end}
F_L(q^2_{\rm max})=1/3 \, , ~~ S_3(q^2_{\rm max})=-1/4 \, , ~~S_4(q^2_{\rm max})=1/4 \, , ~~S_5(q^2_{\rm max})/S_6(q^2_{\rm max})=1/2 \, , ~~S_{5,6,7,8,9}(q^2_{\rm max})=0  \, .
\end{align}

\section{The high-$q^2$ region \label{sec:hiq2}} 

We consider $B\to K^\ast\mu\mu$  decays in the  high-$q^2$ region above the peaking charmonium resonances in the
OPE (Section~\ref{sec:OPE}) and a phenomenological data-driven test case (Section~\ref{sec:KS}).

\subsection{The high-$q^2$ OPE \label{sec:OPE}}

At high $q^2$ one may exploit the presence of this hard scale to employ an OPE~~\cite{Grinstein:2004vb} to control quark-loop effects.
The corresponding contributions   can be absorbed into the effective coefficients of $\mathcal{O}_{7,9}$ following~\cite{Bobeth:2010wg}
\begin{equation}
\begin{split}
\mathcal{C}^{\text{eff}}_7(q^2)&=\mathcal{C}_7-\frac{1}{3}\mathcal{C}_3-\frac{4}{9}\mathcal{C}_4-\frac{20}{3}\mathcal{C}_5-\frac{80}{9}\mathcal{C}_6+\frac{\alpha_s}{4\pi}\bigg[\big(\mathcal{C}_1-6\mathcal{C}_2\big)A(q^2)-\mathcal{C}_8F_8^{(7)}(q^2)\bigg],\\
\mathcal{C}^{\text{eff}}_9(q^2) &=\mathcal{C}_9+\frac{1}{2}h(q^2,0)\bigg[\frac{8}{3}\mathcal{C}_1+2\mathcal{C}_2+11\mathcal{C}_3-\frac{4}{3}\mathcal{C}_4+104\mathcal{C}_5-\frac{64}{3}\mathcal{C}_6\bigg]\\
&+\frac{8}{3}\frac{m_c^2}{q^2}\bigg[\frac{4}{3}\mathcal{C}_1+\mathcal{C}_2+6\mathcal{C}_3+60\mathcal{C}_5\bigg]\\
&+\frac{\alpha_s}{4\pi}\bigg[\mathcal{C}_1\big(B(q^2)+4C(q^2)\big)-3\mathcal{C}_2\big(2B(q^2)-C(q^2)\big)-\mathcal{C}_8F_8^{(9)}(q^2)\bigg]\\
&-\frac{1}{2}h(q^2,m_b^2)\bigg[7\mathcal{C}_3+\frac{4}{3}\mathcal{C}_4+76\mathcal{C}_5+\frac{64}{3}\mathcal{C}_6\bigg]+\frac{4}{3}\bigg[\mathcal{C}_3+\frac{16}{3}\mathcal{C}_5+\frac{16}{9}\mathcal{C}_6\bigg].\label{C9OPE}
\end{split}
\end{equation}
The  functions $F_8^{(7),(9)}$ can be found in~\cite{Beneke:2001at}, while $A,B$ and $C$ are given in~\cite{Seidel:2004jh}. The function $h(q^2,m_q^2)$ specifies the one-loop contributions to the vacuum polarization induced by the quarks and reads
\begin{equation}
  \begin{split}
h(q^2,m_q^2)&=\frac{4}{9}\bigg(\log\frac{\mu^2}{m_q^2}+\frac{2}{3}+w\bigg)-\frac{4}{9}(2+w)\sqrt{\vert w-1\vert}\times
\bigg\{\theta(w-1) \ \displaystyle\arctan \frac{1}{\sqrt{w-1}}\\
&+\theta(1-w)\bigg(
\displaystyle\ln\,\frac{1+\sqrt{1-w}}{\sqrt{z}} - \frac{i\pi}{2}\bigg)\bigg\},\label{h-function}
\end{split}
\end{equation}
with $w=4 m_q^2/q^2$, where $m_q$ denotes the quark's mass. In the limit of the massless quark one finds
\begin{equation}
h(q^2,0)=\frac{8}{27}+\frac{4}{9}\bigg(\log\frac{\mu^2}{q^2}+i\pi \bigg).
\end{equation}
One can then  employ the heavy quark expansion and the operator identities of the QCD to derive the improved Isgur-Wise relations between the (axial)-vector and tensor form factors~\cite{Grinstein:2002cz}. A simple derivation is found in~\cite{Bobeth:2010wg}. After applying these relations one finds that at  leading order in $1/m_b$ the transversity amplitudes are functions of the universal linear combinations of the Wilson coefficients $C^{L,R}$~\cite{Bobeth:2010wg}, namely
\begin{equation}
\begin{split}
A_\perp^{L,R} (q^2)&=+i\,\bigg[\mathcal{C}_9^{\rm eff}(q^2)\mp\mathcal{C}_{10}+\kappa\frac{2 m_b m_B}{q^2}\mathcal{C}_{7}^{\rm eff}(q^2)\bigg]f_\perp(q^2)\equiv +i\,C^{L,R}(q^2) f_\perp(q^2) \, ,\\
A_{0,\parallel}^{L,R}(q^2) &=-i\,\bigg[\mathcal{C}_9^{\rm eff}(q^2)\mp\mathcal{C}_{10}+\kappa\frac{2 m_b m_B}{q^2}\mathcal{C}_{7}^{\rm eff}(q^2)\bigg]f_{0,\parallel}(q^2)\equiv -i\,C^{L,R}(q^2)f_{0,\parallel} (q^2) \,,
\label{amplitudes}
\end{split}
\end{equation}
where  $\kappa=1+(\alpha_s/(3\pi))\ln(m_b^2/\mu^2)$.
Note that the above form of the transversity amplitudes follows from  the universality of  $\mathcal{C}^{\text{eff}}_9(q^2)$, {\it  i.e.}, its independence on the polarization of the final vector meson. This is a property of the high-$q^2$ OPE~\cite{Grinstein:2002cz}. The $1/m_b$-corrections to these relations are parametrically suppressed~\cite{Bobeth:2012vn}. The transversity form factors $f_{\perp,\parallel,0}$ are defined as the following combinations of the standard form factors $A_{1,2}(q^2)$ and $V(q^2)$
\begin{equation}
\begin{aligned}
  \frac{f_{\perp}(q^2)}{N(q^2)} & = \frac{\sqrt{2\, \lambda}}{m_B + m_{K^*}} V(q^2)\,,  \quad \quad 
  \frac{f_{\parallel}(q^2)}{N(q^2)}  = \sqrt{2}\, (m_B + m_{K^*})\, A_1(q^2)\,, \\
\\
  \frac{f_{0}(q^2)}{N(q^2)} & = \frac{(m_B^2 - m_{K^*}^2 - q^2) (m_B + m_{K^*})^2 A_1(q^2)
   - \lambda\, A_2(q^2)}{2\, m_{K^*} (m_B + m_{K^*}) \sqrt{q^2}} = 8 \frac{m_{K^*} m_B}{\sqrt{q^2}} A_{12} (q^2) \, , 
\end{aligned}
\label{eq:FFs}
\end{equation}
\begin{align}
  \label{eq:norm}
  N(q^2) & = G_F\, \alpha_{em}\, V_{tb}^{}V_{ts}^{*}\,
    \sqrt{\frac{q^2 \,  \,\sqrt{\lambda}}{3 \cdot 2^{10}\, \pi^5\, m_B^3}} \, , 
    \end{align}
where $\lambda=\lambda(q^2, m_B^2, m_{K^*}^2)$ denotes the K\" all\' en function $\lambda(a,b,c)=a^2+b^2+c^2-2(a b+a c+b c)$.
The form factor $f_0(q^2)$ is proportional to $A_{12}(q^2)$ that has been directly computed in  Lattice QCD~\cite{Horgan:2013hoa}. 

To ease notation in the remainder of this work we frequently drop the explicit $q^2$-dependence for transversity amplitudes, form factors, effective coefficients etc.

\subsection{The Kr\"uger-Sehgal approach  \label{sec:KS}} 

In this section we describe the method which aims at a local description of charm resonances in the high-$q^2$ region of $B\to K^{(\ast)}\ell\ell$ decays using the charm contribution to the self-energy of the 
photon and a factorization ansatz.
 The full non-perturbative form of the charm vacuum polarization function can be extracted from data on the $e^+e^-\to h_i$ scattering, where $h_i$ denotes all produced hadrons in the given kinematic region. This idea was first proposed in~\cite{Lim:1988yu, Kruger:1996cv}. Such analysis was recently performed in Ref.~\cite{Lyon:2014hpa} for the case $B\to K\mu\mu$.

The experimentally accessible observable is the ratio of the cross section of $e^+e^-$ scattering into hadrons normalized to the corresponding cross section of the scattering into muon pairs as the function of the center-of-mass energy $s\equiv q^2$, namely
\begin{equation}
R(s)=\frac{\sigma^{(e^+e^-\to h_i)}(s)}{\sigma^{(e^+e^-\to \mu^+\mu^-)}(s)}.\label{Rfit}
\end{equation}
We fit for the function $R(s)$ in the interval $\sqrt{s}=3.7\,\mbox{GeV}$ to $\sqrt{s}=4.8\,\mbox{GeV}$ using the available data on the $e^+e^-\to h_i$ processes from the BES experiment~\cite{Bai:2001ct, Ablikim:2007gd}. The ratio $R(s)$ is the sum of the resonant and the continuum contributions
\begin{equation}
R(s)=R_{\text{res}}(s)+R_{\text{cont}}(s).\label{Rfit2}
\end{equation}
The explicit form of  $R_{\text{res, cont}}(s)$ with  further details of the fitting procedure can be found in Appendix~\ref{Details of the fit for $R(q^2)$}.
The charm contribution to Eq.~\eqref{Rfit} is extracted using 
\begin{equation}
R_c(s)=R(s)-R_{uds},\label{Rc}
\end{equation} 
where $R_{uds}=2.16$ is the asymptotic value of the light-quark contributions.

The relevant scattering amplitude can be written as 
\begin{equation}
\mathcal{A}{(e^+e^-\to h_{(c\bar{c})}\to e^+e^-)}=\frac{e^4}{s^2}(\bar{e}\gamma_\mu e)(\bar{e}\gamma_\nu e)\Pi^{\mu\nu(c)}(s).
\end{equation}
Gauge invariance  dictates the form of the photon's self-energy, $\Pi_{\mu\nu}(s)=(-g_{\mu\nu}q^2+q_\mu q_\nu)\Pi(s)$.
The optical theorem relates the imaginary part of this amplitude to the total hadronic cross section, which implies
\begin{equation}
R_c(s) =\frac{e^2 \operatorname{Im}[\Pi^{(c)} (s)] }{e^2 \operatorname{Im}[\Pi^{(\mu)}(s)  ]} \, .
\end{equation}
For easier comparison with Kr\"uger and Sehgal (KS) \cite{Kruger:1996cv}, we introduce
\begin{equation}
\Pi^{(KS)}(s)\equiv e^2 \operatorname{Im}[\Pi^{(c)}(s) ] \, .
\end{equation}
Using $\operatorname{Im}[\Pi^{(\mu)}]=1/(12\pi)$ we obtain
\begin{equation}
\operatorname{Im}[\Pi^{(KS)}(s)]=\frac{\alpha_{\text{em}}}{3}R_c(s).\label{PiKS}
\end{equation}
The charm polarization function $h_c(s)$ is  defined in such a way to match the perturbative evaluations in Eq.~\eqref{h-function}, as
\begin{equation}
h_c(q^2)=\frac{\pi}{\alpha_{\text{em}}}\Pi^{(KS)}(q^2)   \, . \label{hc}
\end{equation}
Then Eqs. \eqref{hc} and \eqref{PiKS} imply
\begin{equation}
\operatorname{Im}[h_c(q^2)]=\frac{\pi}{3}R_c(q^2)  \, .
\end{equation}
Together with Eq.~\eqref{Rc} we extract the imaginary part of the function $h_c(q^2)$ from the fit for the function $R(s)$.
\begin{figure}[H]
\subfigure{\includegraphics[width=0.48\textwidth]{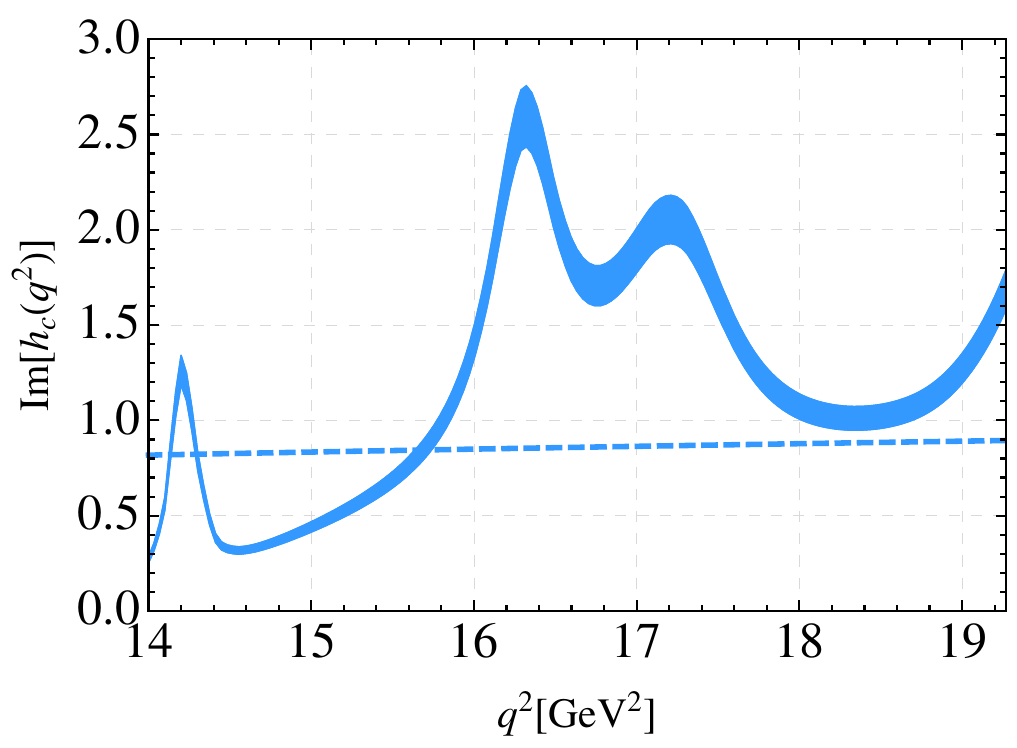}}
\subfigure{\includegraphics[width=0.497\textwidth]{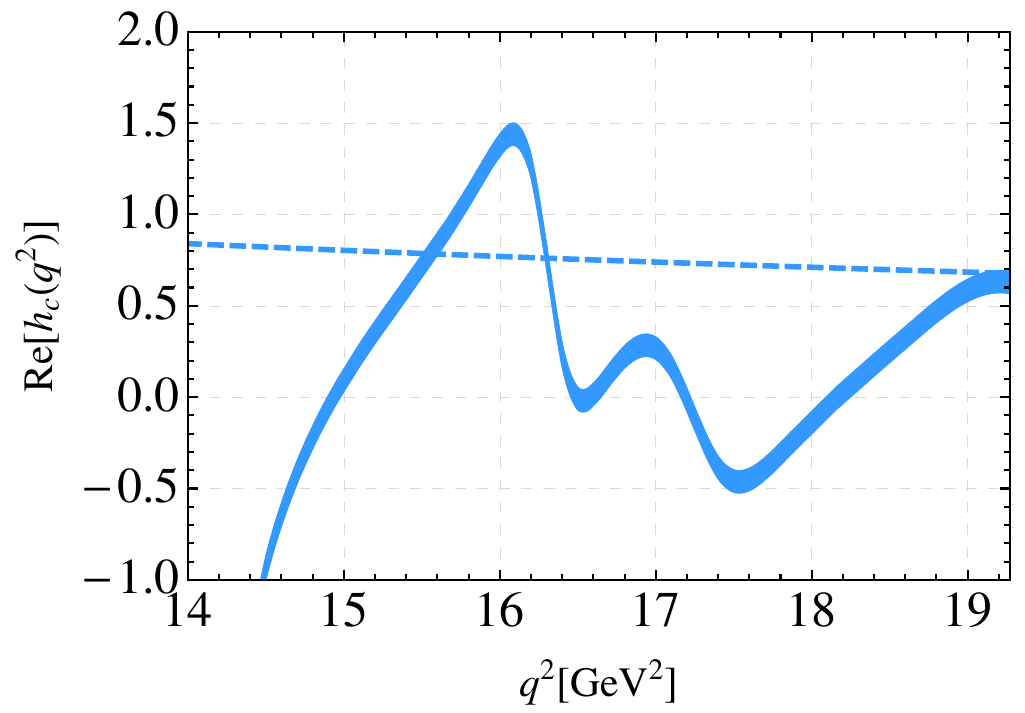}}
\caption{The imaginary and the real part of the charm polarization function $h_c(q^2)$ extracted from the fit (blue $1 \sigma$ band)  to $e^+e^-\to h_i$ data from BES-II \cite{Bai:2001ct}  in the region $q^2 {\in} (3.6^2,4.8^2)\,\mbox{GeV}^2$. The  corresponding OPE-contributions,  imaginary and  real part of  $h(q^2,m_c^2)$, are shown by the blue dashed lines.} \label{Fig.1}
\end{figure}
We obtain the real part of $h_c(s)$ from its imaginary part  using the subtracted dispersion relation
\begin{equation}
\operatorname{Re}[h_c(s)]=\operatorname{Re}[h_c(s_0)]+\frac{s-s_0}{\pi}P\int_{t_0}^{\infty}\! \frac{ds'}{(s'-s)(s'-s_0)}\operatorname{Im}[h_c(s')],\label{dispersion relation}
\end{equation}
where the arbitrary subtraction point $s_0$ and the lower limit of integration $t_0$ are convincingly chosen in the perturbative regime below the $J/\psi$-resonance peak and $P$ denotes the principal part.
The function $h_c$ is shown in Fig.~\ref{Fig.1}.

We proceed using a factorization ansatz and absorb $\Pi^{(KS)}$ into the charm contribution of the effective coefficient of $\mathcal{O}_9$. 
The corresponding $B \to K^* \ell  \ell $ matrix element, which includes $B \to K^* (\bar c c) \to K^*  \ell \ell $  charmonium contributions,
can be obtained by replacing the propagating resonances with  the self-energy $\Pi^{\mu\nu}(q^2)$
\begin{equation}
\mathcal{M}=- \frac{G_F}{2 \sqrt{2}} V_{tb}^{\vphantom{*}} V_{ts}^*  3 a_2^{\vphantom{*}} \eta_c^{\vphantom{*}} \Pi^{(KS)}(q^2)
\langle K^{*} \vert\bar{s}\gamma_\mu(1-\gamma_5) b\vert B\rangle \, \bar{\ell}\gamma^\mu\ell\  \, .    \label{LD}
\end{equation}
Therefore,
\begin{equation}
  \begin{split}
\mathcal{C}_9^{\rm eff}(q^2)&=\mathcal{C}_9+ 3 a_2 \, \eta_c \, h_c(q^2)-\frac{1}{2}h(q^2,0)\bigg[\mathcal{C}_3+\frac{4}{3}\mathcal{C}_4+16\mathcal{C}_5+\frac{64}{3}\mathcal{C}_6\bigg]\\
&-\frac{1}{2}h(q^2,m_b^2)\bigg[7\mathcal{C}_3+\frac{4}{3}\mathcal{C}_4+76\mathcal{C}_5+\frac{64}{3}\mathcal{C}_6\bigg]+\frac{4}{3}\bigg[\mathcal{C}_3+\frac{16}{3}\mathcal{C}_5+\frac{16}{9}\mathcal{C}_6\bigg].
\end{split}
\label{C9effFA}
\end{equation}
Here, we explicitly included  terms that arise from the perturbative $b$- and light-quark contributions. 
$a_2$ is a combination of Wilson coefficients that accounts for the perturbative charm-loop
\begin{equation} \label{eq:a2}
a_2=\frac{1}{3}\bigg(\frac{4}{3}\mathcal{C}_1+\mathcal{C}_2+6 \mathcal{C}_3 + 60 \mathcal{C}_5\bigg) \, .
\end{equation}
To be specific, in this work we employ  the value obtained at next-to-next-to-leading order (NNLO) at the $b$-mass scale, $a_2=0.2$ in the numerical analyses. (In the operator basis used in earlier works $3 a_2$ corresponds to $C^{(0)}$~\cite{Ali:1999mm}.)
Furthermore, we introduced in Eq.~(\ref{C9effFA}) a  fudge function  $\eta_c \equiv \eta_c(K^*_j,q^2)$ that corrects for effects beyond factorization. In general $\eta_c$  is complex-valued and depends on the transversity state of the $K^*$, $j=\perp, \parallel,0$.
For previous usage of fudge factors, see, {\it e.g.},~\cite{Ali:1991is, Ali:1999mm, Ligeti:1995yz, Lyon:2014hpa,Khodjamirian:2010vf}.
Note,  that in principle, a dependence on  the decay angles is possible as well: $\theta_\ell$-dependence can arise from electromagnetic corrections, while $\theta_K$-dependence can arise from the $K^\ast$ beyond the narrow width approximation, however, both of these effects are neglected in this work.

\section{Wiggles and non-universality}
\label{sec:wiggles}
\begin{figure}[]
\begin{center}
\subfigure{\includegraphics[width=0.40\textwidth]{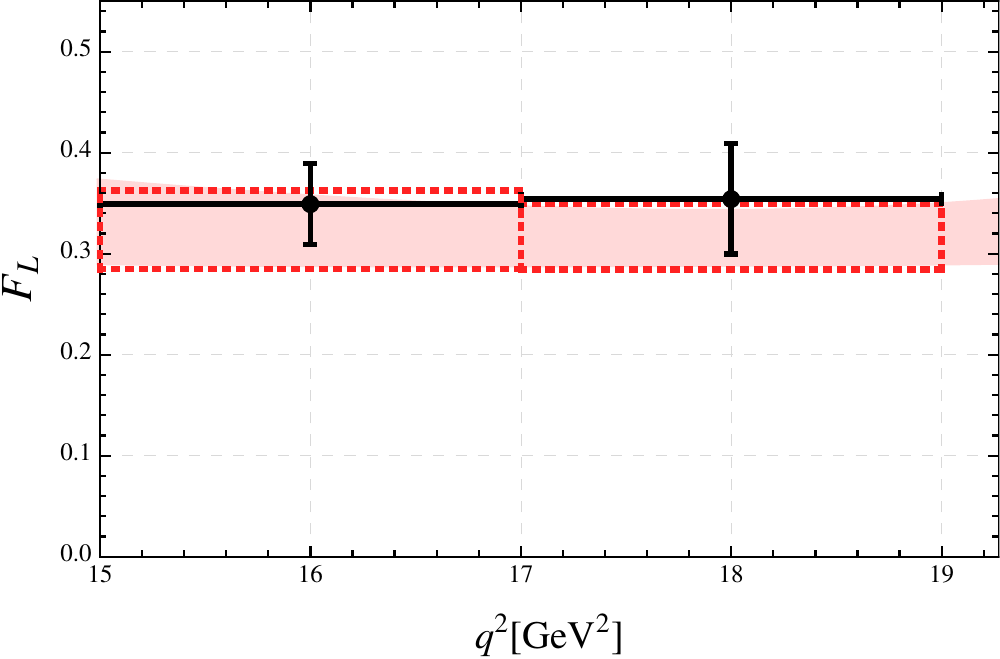}}
\subfigure{\includegraphics[width=0.40\textwidth]{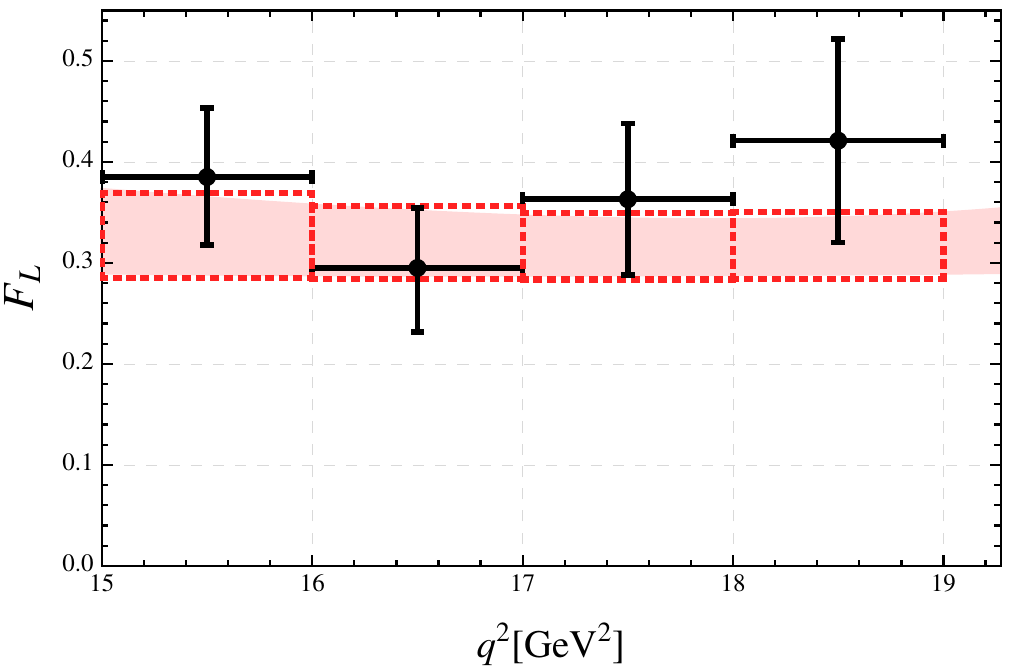}}
\subfigure{\includegraphics[width=0.40\textwidth]{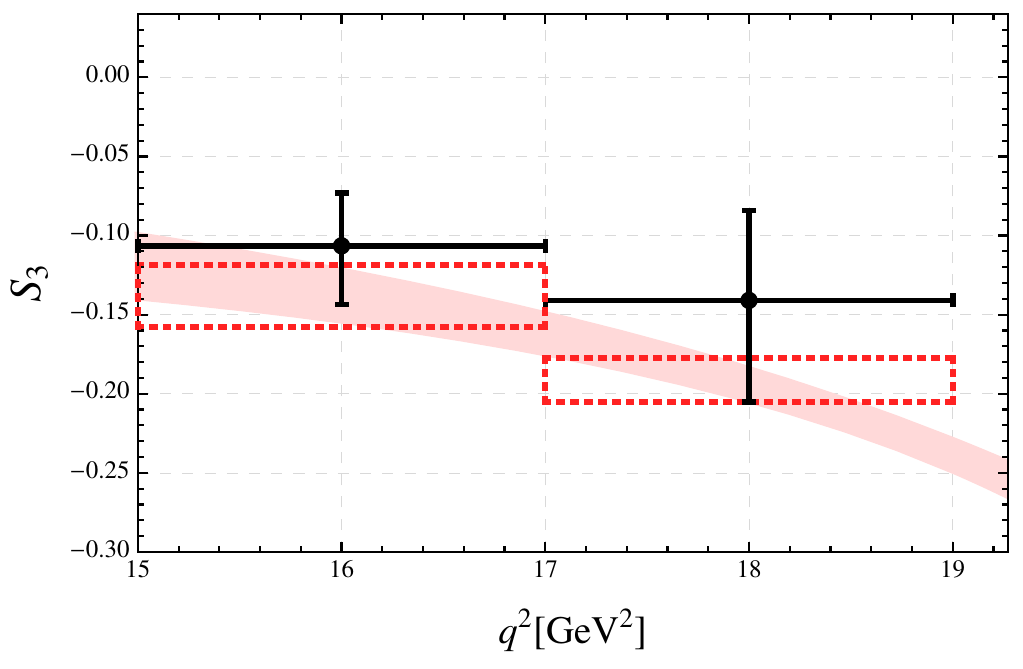}}
\subfigure{\includegraphics[width=0.40\textwidth]{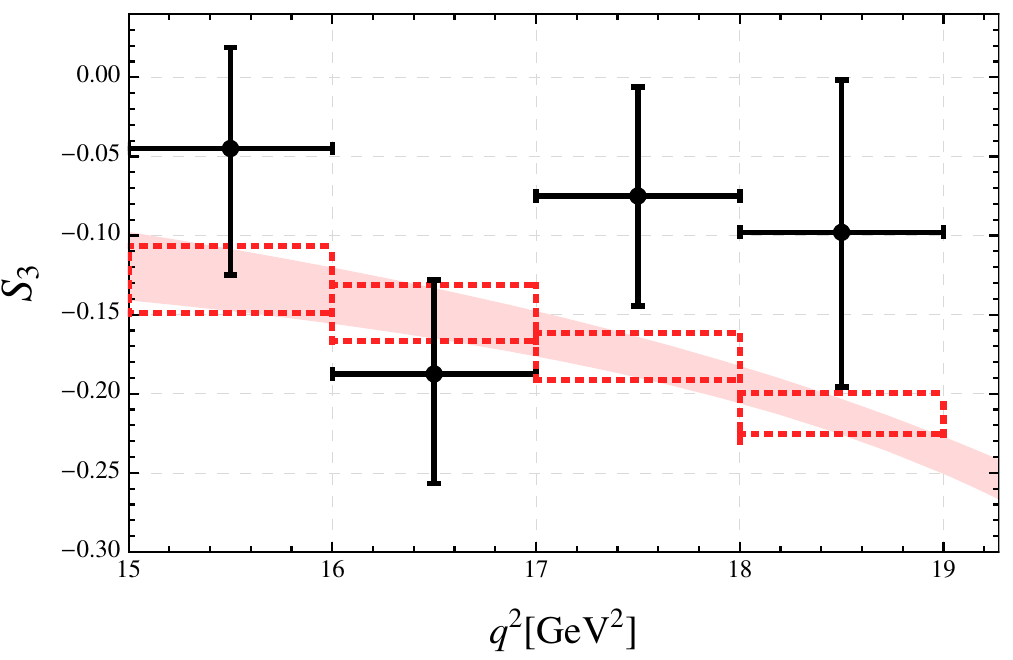}}
\subfigure{\includegraphics[width=0.40\textwidth]{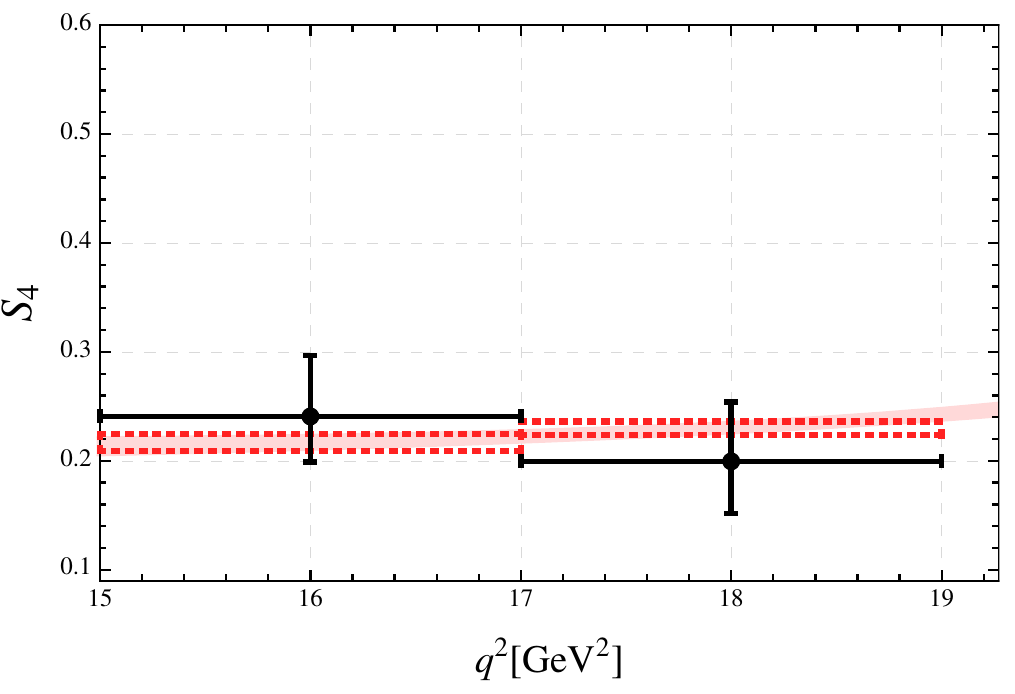}}
\subfigure{\includegraphics[width=0.40\textwidth]{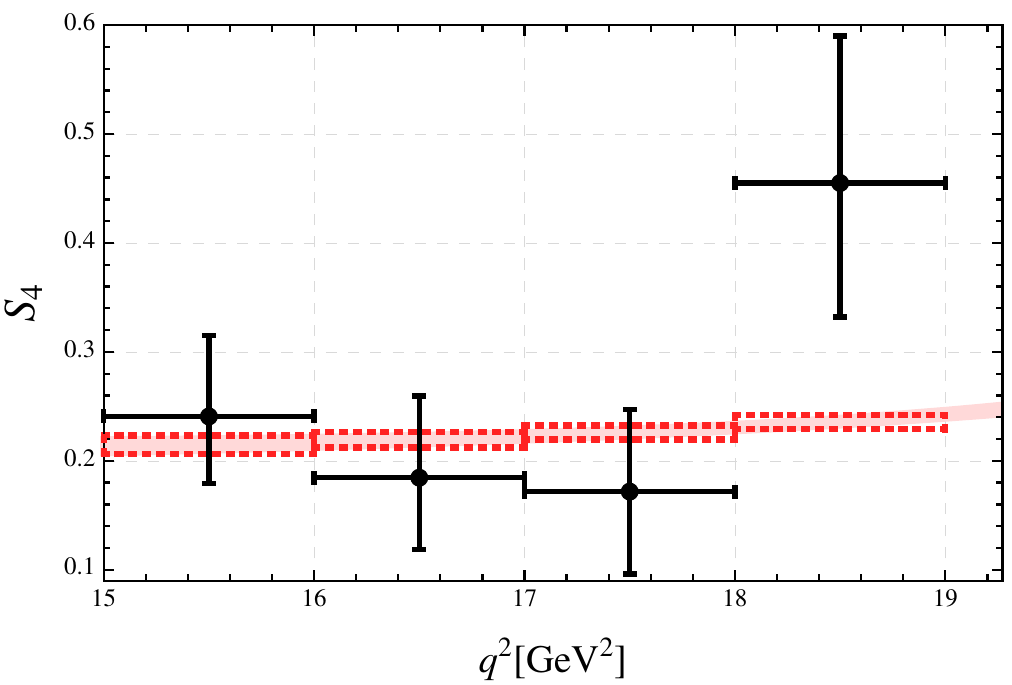}}
\caption{The angular observables  $F_L, S_{3}$ and $S_4$ in the OPE for 2 $\mbox{GeV}^2$ bins (plots to the left) and 1 $\mbox{GeV}^2$ bins (plots to the right) 
shown as red boxes versus data (black) from LHCb \cite{Aaij:2015oid}. Systematic and statistical uncertainties are added in quadrature.
The light-shaded red bands illustrate the OPE for infinitesimal binning. Form factors are taken from \cite{Horgan:2013hoa}. 
The binned observables approach the continuous functions in the limit of infinitesimal bin width. }
 \label{PlotsFLcont}
\end{center}
\end{figure}

Both wiggles in binned $q^2$-distributions and non-universality would signal a breakdown of the OPE. We  compare the predictions of the OPE (red curves and boxes with form factors from~\cite{Horgan:2013hoa}) to data (black) in
Figs.~\ref{PlotsFLcont} and~\ref{fig:S5AFB}, zooming in from 2 $\mbox{GeV}^2$ bins (plots to the left) to finer resolution with 1 $\mbox{GeV}^2$ bins (plots to the right).
 Quite generally  one expects an onset of resonance structure, consistent with the measured $R$-ratio \cite{Ablikim:2007gd}, see also Figs.~\ref{Fig.1}
and \ref{fig:Rfit}. As the branching ratio has not been measured with resolution smaller than 2 $\mbox{GeV}^2$ bins we only show this  binning in Fig.~\ref{fig:Br}.
From  these figures  one cannot draw  firm conclusions on observing a resonance structure in any of the observables due to the limited experimental precision. While  currently resonance effects  are not noticeable  in the $2 \, \mbox{GeV}^2$ bins, the alternating patterns in the 1 $\mbox{GeV}^2$ bins, however, may be hinting at such as structure. 
Further data with improved precision is required to clarify this point.

In addition, the data have to meet 
 the endpoint predictions (\ref{eq:end}) irrespective of BSM contributions.
 A significant violation of Eq.~(\ref{eq:end})  would, for instance, point to underestimated backgrounds other than from $K^* \to K \pi$.
 In particular with 1 $\mbox{GeV}^2$ bins  data on $S_{3,4,5}$ are  presently in mild conflict  at $\sim 1-2 \sigma$ with the endpoint relations.
 Note, however, that the  endpoint bin is challenged by the dying statistics and needs to be viewed with a grain of salt, see Fig.~\ref{PlotsFLcont}.

In Section \ref{sec:freedom} we discuss different classes of angular observables according to their sensitivity to short-distance physics and resonance parameters.
As wiggles and non-universality are both effects beyond the OPE, yet need to be measured, therefore, and only therefore, we use the phenomenological KS-approach as an 
efficient parameterization of local spectra. 
In Section \ref{sec:uni} we work out phenomenological constraints on the resonance parameters.

\subsection{Short-distance freedom and short-distance  sensitivity \label{sec:freedom}} 

The universal feature of the OPE-amplitudes \eqref{amplitudes} enables the construction of observables in which the dependence on the short-distance  coefficients, $C^{L,R}$, cancels  \cite{Bobeth:2010wg}. $F_L, S_3,S_4$ belong to this class of short-distance free observables, which
are defined in terms of transversity amplitudes as follows, respectively,
\begin{equation}
\small
F_L\equiv \frac{\vert A_0^L\vert^2+\vert A_0^R\vert^2}{d\Gamma/dq^2},\,\,
S_3=\frac{3}{8}\frac{\vert A_\perp^L\vert^2-\vert A_\parallel ^L\vert^2+\vert A_\perp^R\vert^2-\vert A_\parallel^R\vert^2}{d\Gamma/dq^2},\,\, S_4=\frac{3}{4 \sqrt{2}}\frac{Re(A_0^LA_\parallel^{L\ast}+A_0^RA_\parallel^{R\ast})}{d\Gamma/dq^2} \, , 
\label{SDfree-observables}
\end{equation}
where the differential decay rate is given as
\begin{equation} \label{eq:br}
\frac{d\Gamma}{dq^2}=\sum_{j=0,\parallel,\perp} \frac{d\Gamma_j}{dq^2} \, , \quad  \frac{d\Gamma_j}{dq^2} =\vert A_j^L\vert^2+\vert A_j^R\vert^2 \, .
\end{equation}
Inserting the transversity amplitudes \eqref{amplitudes} into Eq.~\eqref{SDfree-observables}, one finds that  the dependence on Wilson coefficients in the limit of the infinitesimal bin width cancels
\begin{equation}
\begin{split}
F_L=\frac{f_0^2}{f_0^2+f_\parallel^2+f_\perp^2} \, ,\quad
S_3=\frac{3}{8}\frac{f_\perp^2-f_\parallel^2}{f_0^2+f_\parallel^2+f_\perp^2}  \, , \quad
S_4=\frac{3}{4\sqrt{2}}\frac{f_0 f_\parallel}{f_0^2+f_\parallel^2+f_\perp^2}   \, .   \label{nice-eq}
\end{split}
\end{equation}

The measured observables correspond to the binned values of the angular coefficients, {\it i.e.}, $\langle J_i\rangle=\int_{bin} J_i dq^2$ such that  in \eqref{nice-eq} products 
of type $f_i f_j$, for $i=0,\parallel,\perp$ are integrated as $\int_{bin} \rho_1f_if_j dq^2$, where 
\begin{equation}
\begin{split} \label{eq:ro1}
\rho_1(q^2)\equiv \frac{1}{2}(\vert C^R\vert^2+ \vert C^L\vert^2)=\bigg\vert\mathcal{C}_9^{\rm eff}+\kappa\frac{2 m_b m_B}{q^2}\mathcal{C}_7^{\rm eff}\bigg\vert^2+\vert\mathcal{C}_{10}\vert^2 \, .
\end{split}
\end{equation}
Since the effective coefficients  that follow from the OPE are slowly varying functions of  $q^2$ the resulting binning effect is small~\cite{Hambrock:2012dg}.

By the same argument which makes $F_L, S_3,S_4$  short-distance insensitive  contributions to $\mathcal{C}_9^{\rm eff}$ as in Eq.~(\ref{C9effFA}) with universal $\eta_c(K^*_j,q^2)$ drop out in  these observables.
The good agreement between the data and the OPE shown in Fig.~\ref{PlotsFLcont} therefore implies that there are no extremely large contributions from non-universal pieces.
It also implies constraints on right-handed currents, which could spoil Eq.~\eqref{nice-eq}.\footnote{In the presence of chirality-flipped operators beyond  (\ref{EffHamiltonian2}) the apparent universality of the short-distance coefficients following from the lowest order OPE, Eq.~(\ref{amplitudes}), breaks down to a partial one. Specifically, only  the longitudinal and parallel coefficients remain the same.}
Such BSM effects, however, would  induce  shapes essentially flat in $q^2$. 
\begin{figure}[]
\begin{center}
\subfigure{\includegraphics[width=0.40\textwidth]{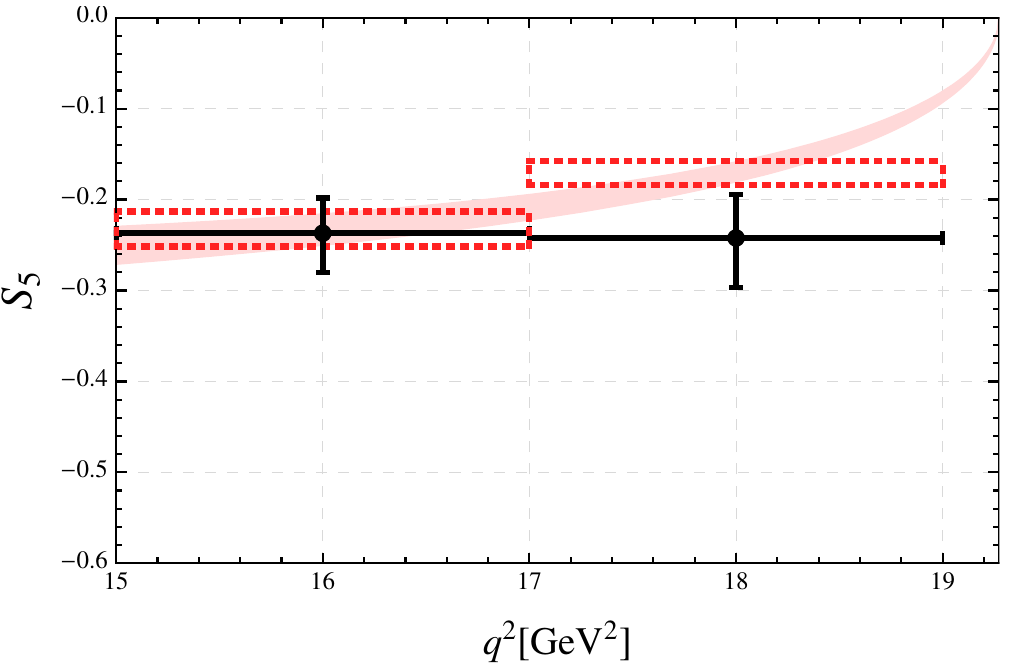}}
\subfigure{\includegraphics[width=0.40\textwidth]{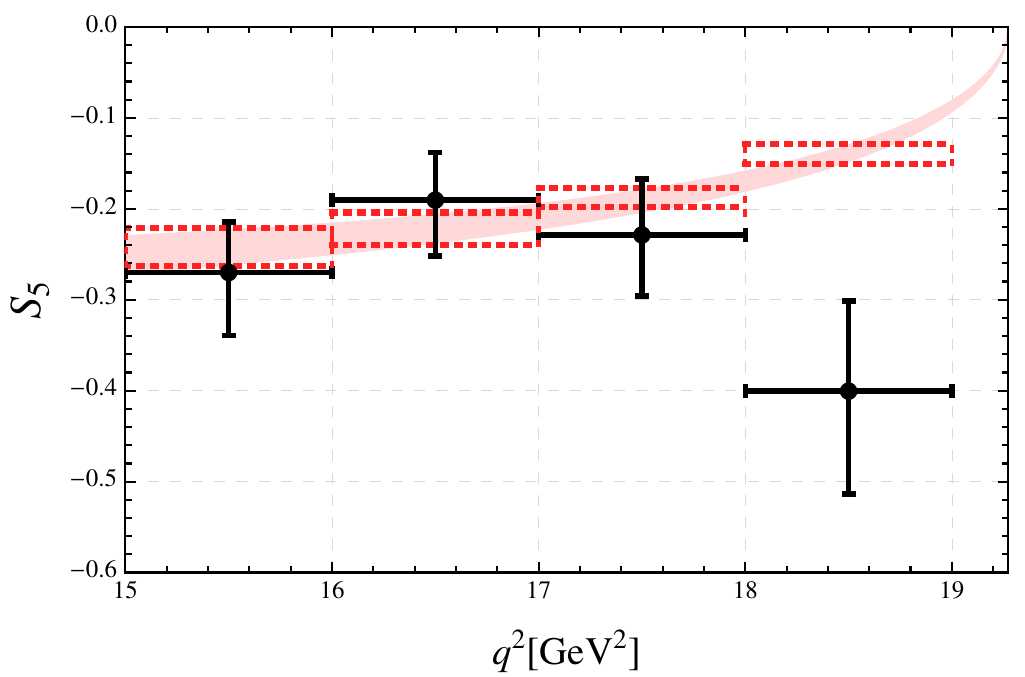}}
\subfigure{\includegraphics[width=0.40\textwidth]{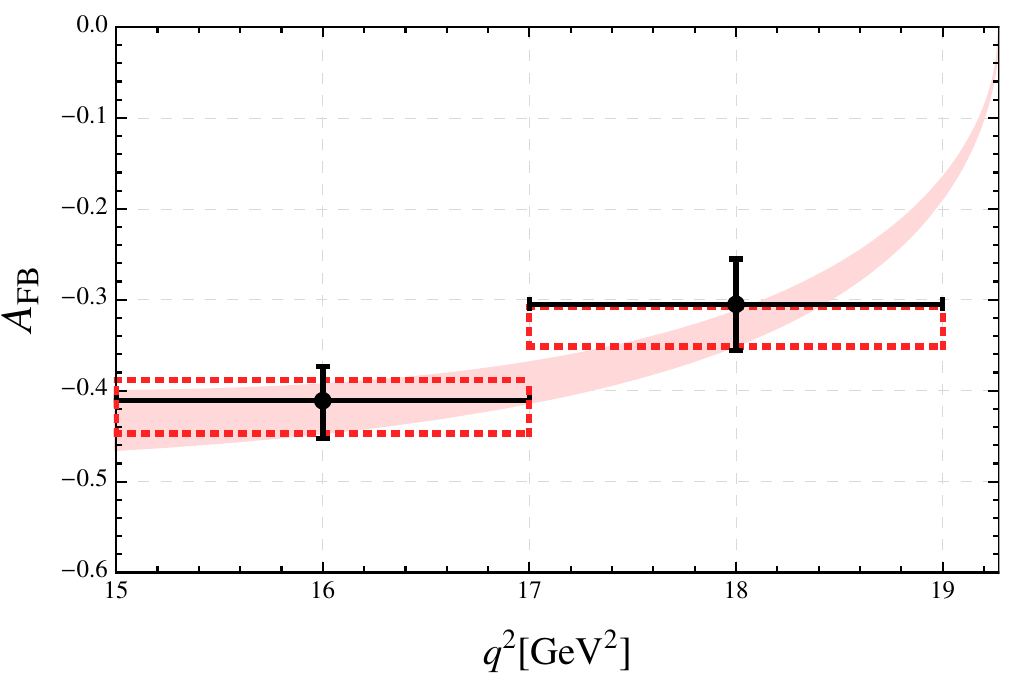}}
\subfigure{\includegraphics[width=0.40\textwidth]{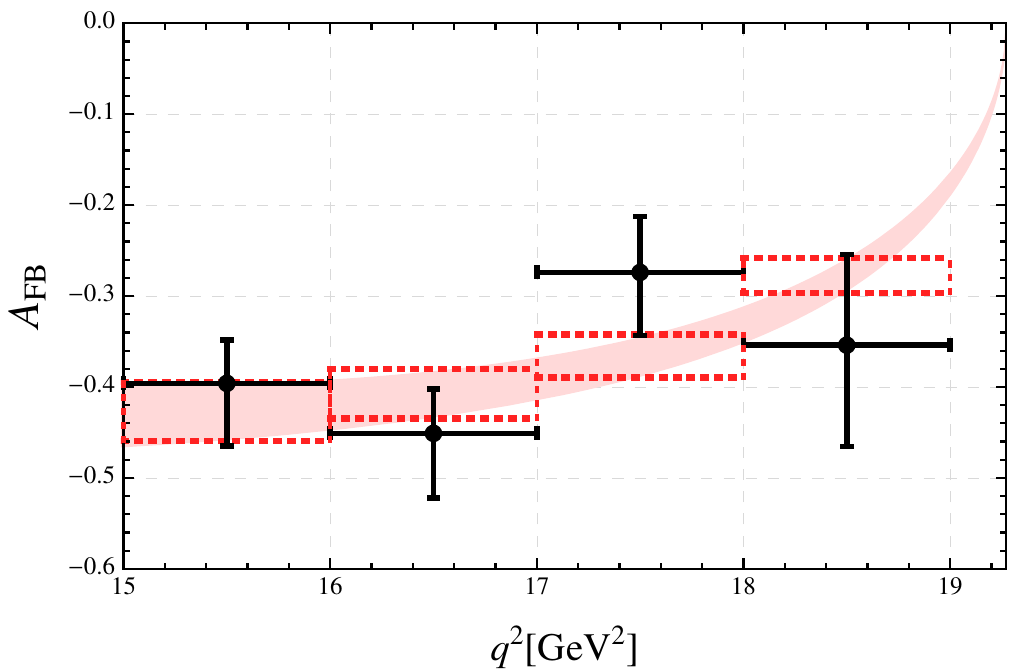}}
	\caption{The  angular observables $S_5$ and  $A_{\rm FB}$ in the OPE in the SM (red boxes) versus data (black)  \cite{Aaij:2015oid} as in Fig.~\ref{PlotsFLcont}.}
\label{fig:S5AFB}
\end{center}
\end{figure}
Another class of observables are short-distance dependent angular observables. These include $A_{\rm FB}, S_6$ and $S_{7,8,9}$.
The former  read in terms of transversity amplitudes
\begin{equation}
S_5=\frac{3\sqrt{2}}{4}\frac{\operatorname{Re}(A_0^LA_\perp^{L\ast}-A_0^RA_\perp^{R\ast})}{d\Gamma/dq^2} \, , \quad
A_{\rm FB}\equiv S_{6}=\frac{3}{4}\frac{2\operatorname{Re}(A_\parallel^LA_\perp^{L\ast}-A_\parallel^RA_\perp^{R\ast})}{d\Gamma/dq^2} \, .
\end{equation}
Within the OPE~\eqref{amplitudes} these observables reduce to
\begin{equation}
\begin{split}
S_5&=\frac{3\sqrt{2}}{2}\frac{\rho_2(q^2)f_0f_\perp}{\rho_1(q^2)\big(f_0^2+f_\perp^2+f_\parallel^2\big)} \, , \quad
A_{\rm FB}=\frac{3\rho_2(q^2)f_\parallel f_\perp}{\rho_1(q^2)\big(f_0^2+f_\perp^2+f_\parallel^2\big)} \, , \label{nice-eq-2}
\end{split}
\end{equation}
with 
\begin{equation}
\begin{split} \label{eq:ro2}
\rho_2(q^2)\equiv \frac{1}{4}(\vert C^R\vert^2- \vert C^L\vert^2)={\rm Re}\bigg[\bigg(\mathcal{C}_9^{\rm eff}+\kappa\frac{2 m_b m_B}{q^2}\mathcal{C}_7^{\rm eff}\bigg)\mathcal{C}^\ast_{10}\bigg] \, ,
\end{split}
\end{equation}
and $\rho_1$ can be seen in Eq.~(\ref{eq:ro1}).
$A_{\rm FB}$ and $S_6$ are shown in the SM in Fig.~\ref{fig:S5AFB}, where we employ the SM Wilson coefficients \eqref{EffHamiltonian1}, \eqref{EffHamiltonian2}, evaluated at  NNLO~\cite{Gorbahn:2004my, Bobeth:1999mk}.
Universality predicts further  $J_{7,8,9}=0$ \cite{Bobeth:2010wg}, and consequently
\begin{align}
S_{7,8,9}=0 \, ,
\end{align}
which can be explicitly seen from Appendix~\ref{sec:Ji}.

\begin{figure}[]
\begin{center}
\subfigure{\includegraphics[width=0.40\textwidth]{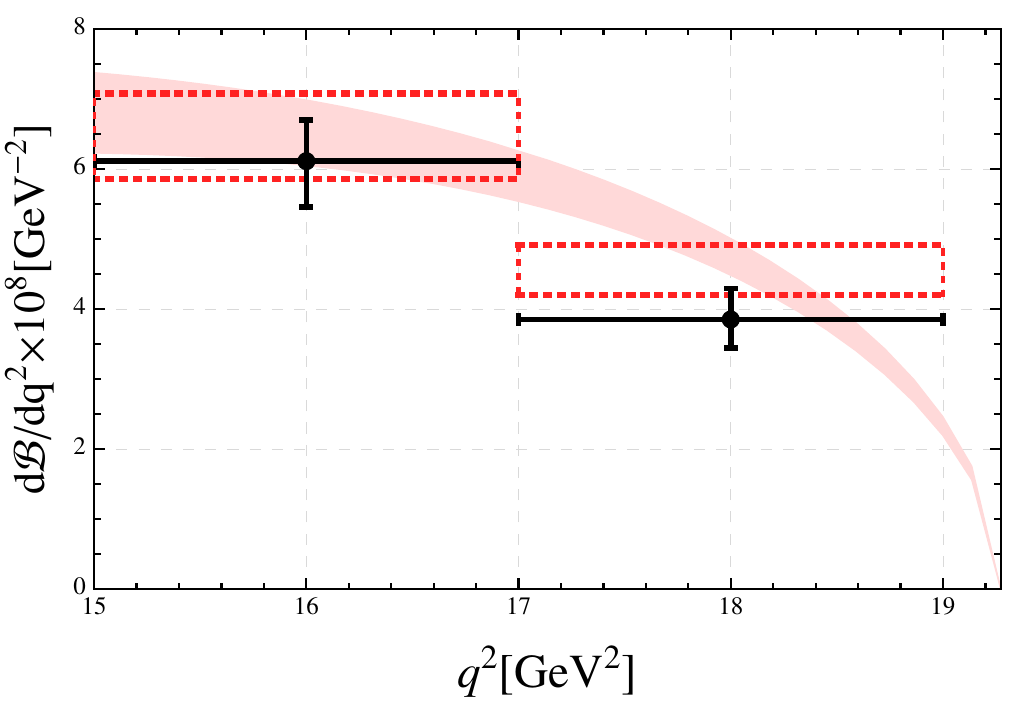}}
\caption{The dilepton invariant mass spectrum $d{\cal{B}}/dq^2$ in the OPE in the SM versus data (black) \cite{LHCb-BR}, see  Fig.~\ref{PlotsFLcont}.}\label{fig:Br}
\end{center}
\end{figure}

The branching ratio, shown in Fig.~\ref{fig:Br} for  the smallest available binning, depends on  BSM physics and is highly sensitive to wiggles whether universal or not,  as no cancellations as in the previously discussed observables
can take place. Also in the branching ratio  the OPE plus SM is in agreement within $1 \sigma$ with the data \cite{LHCb-BR}.
We learn that in order to maximally probe for local structures and their deviations from binned OPE-results one has to simultaneously fit to BSM coefficients and resonance parameters.

\subsection{Probing resonances}
\label{sec:uni}

In this section we extract information from data on the $\eta_c(K^*_i, q^2)$-parameters.
To begin we note that by means of Lorentz invariance all non-factorizable contributions have to vanish at the endpoint $q^2_{\rm max}$ \cite{Hiller:2013cza}.
This implies 
\begin{align} \label{eq:EP}
\eta_c(K^*_0, q^2_{\rm max})=\eta_c(K^*_\parallel, q^2_{\rm max}) \, .
\end{align}
To facilitate a fit already with presently available data we assume constant $\eta_c$-functions in the entire high-$q^2$ region. This is, of course, a simplifying working-assumption,
however, as we show in Section \ref{sec:SM}, it describes $B \to K^* \mu \mu$ data  well. Similarly,
there is support for this  from $B \to K \mu \mu$ data, which also gives a good fit for this assumption~\cite{Lyon:2014hpa}. With better data one should investigate more general shapes.

Let us illustrate how $S_{7,8,9}$ are informative for non-universal $\eta_c$ \cite{Bobeth:2012vn}, as within our assumptions, approximately,
\begin{align}
J_7  &\simeq  -\frac{9}{\sqrt{2}}  f_0 f_\parallel  \mathcal{C}_{10} a_2 {\rm  Im} [ h_c(q^2) (\eta_c(K^*_0,q^2)- \eta_c(K^*_\parallel,q^2)) ] \, , \\
J_8  &
\simeq - \frac{9}{2 \sqrt{2}}  f_0 f_\perp (  \mathcal{\tilde C}_9^{\rm eff}+\kappa\frac{2 m_b m_B}{q^2}\mathcal{C}_{7}^{\rm eff})  a_2 {\rm Im } [h_c (q^2) (\eta_c(K^*_0,q^2)- \eta_c(K^*_\perp,q^2)) ]\, , \\
J_9 &  \simeq  -\frac{9}{2} f_\parallel f_\perp (   \mathcal{ \tilde C}_9^{\rm eff}+\kappa\frac{2 m_b m_B}{q^2}\mathcal{C}_{7}^{\rm eff})  a_2 {\rm  Im} [h_c (q^2)(\eta_c(K^*_\parallel,q^2)- \eta_c(K^*_\perp,q^2))]  \, ,
\end{align}
where $\tilde C_9^{\rm eff}$ equals $C_9^{\rm eff}$ with the charm contribution removed. Terms quadratic in $a_2$ have not been spelled out explicitly; they require relative phases in the $\eta_c$ and are  mildly suppressed by $3 a_2/C_9$.

\begin{figure}[]
\begin{center}
\subfigure{\includegraphics[width=0.37\textwidth]{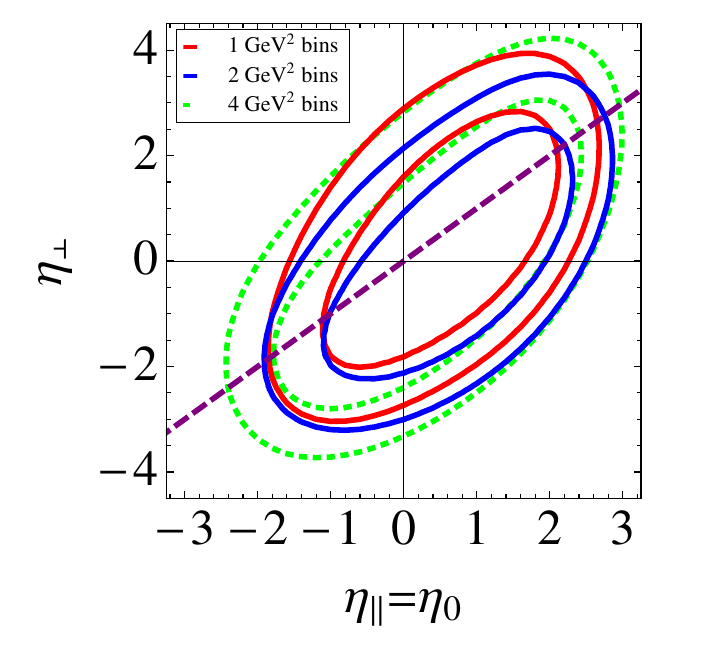}}
\subfigure{\includegraphics[width=0.33\textwidth]{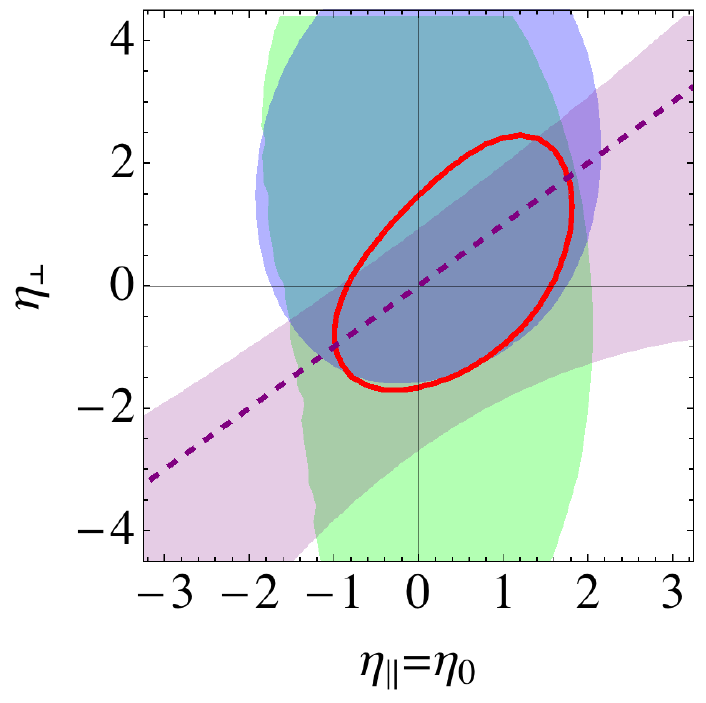}}
\caption{Left plot: Allowed 68 \% C.L. (inner) and 95 \% C.L. (outer) regions from a simultaneous fit to $\delta \mathcal{C}_9, \delta \mathcal{C}_{10}$ and $\eta_\parallel=\eta_0$ and $\eta_\perp$ using 4 $\mbox{GeV}^{2}$ bins (green contours), 2 $\mbox{GeV}^{2}$ bins (blue contours) and 1 $\mbox{GeV}^{2}$ bins (red contours), see text. The dashed magenta line denotes the universality limit $\eta_\perp=\eta_\parallel=\eta_0$.
  Right plot:  Illustration of allowed 68 \% C.L (inner) regions for $\eta_\parallel=\eta_0$ and $\eta_\perp$ in the SM using
 1 $\mbox{GeV}^{2}$ bins for $S_8,S_9$ (magenta area), $A_{\rm FB}$ (blue area) and 2 $\mbox{GeV}^{2}$ bins for the branching fraction~\cite{LHCb-BR} (green area).}
\label{fig:NU-fit}
\end{center}
\end{figure}

In order to comply with Eq.~(\ref{eq:EP}) we fix $\eta_0=\eta_\parallel$\footnote{Data on  $B\to K^* ( \psi, \psi(2S))$, far away from the endpoint, indicate indeed  $0.8 \lesssim \vert \frac{\eta_\parallel}{\eta_0} \vert \lesssim 1.4$, see Appendix~\ref{sec:Kst} for details.}
and fit simultaneously to the resonance parameters $\eta_0,\eta_\perp$ and the BSM Wilson coefficients $\delta \mathcal{C}_9, \delta \mathcal{C}_{10}$, using all three available binnings. Here and in the following we abbreviate $\eta_j \equiv \eta_c(K^*_j,q^2)$, $j=0, \parallel, \perp$. 
In Fig.~\ref{fig:NU-fit} (plot to the left) we show constraints on $\eta_0=\eta_\parallel$  and $\eta_\perp$ for 4 $\mbox{GeV}^2$ bins (green contours), 2 $\mbox{GeV}^2$ bins (blue contours) and 1 $\mbox{GeV}^2$ bins (red contours).
Naive factorization $\eta_j=1$ is allowed, but also values away from universality, the latter indicated by the dashed magenta line.
The fits are also consistent with no charm-loop contribution, $\eta_j=0$.
However, modulo experimental effects, the small binning-induced differences  between constraints may hint at the presence of such structure.

The weakest constraints stem from the largest bin size. We stress that the constraints in the left plot of Fig.~\ref{fig:NU-fit} are obtained without assuming the SM.
The corresponding predictions of the fit for the BSM coefficients are discussed in the next section~\ref{sec:SM}.
For illustrational purposes we show  in the plot to the right of Fig.~\ref{fig:NU-fit} the allowed $1 \sigma $ contour for $\eta_0=\eta_\parallel$ and $\eta_\perp$ for 1 $\mbox{GeV}^{2}$ bins in the SM.
The constraints are more tight  than in the model-independent fit.
We also show  the individual 1$\sigma$ areas of various constraining observables,
$S_8,S_9$ (magenta area),  $A_{\rm FB}$ (blue area) and 2 $\mbox{GeV}^{2}$ bins for the branching fraction~\cite{LHCb-BR} (green area). 

\begin{figure}[]
\begin{center}
\subfigure{\includegraphics[width=0.40\textwidth]{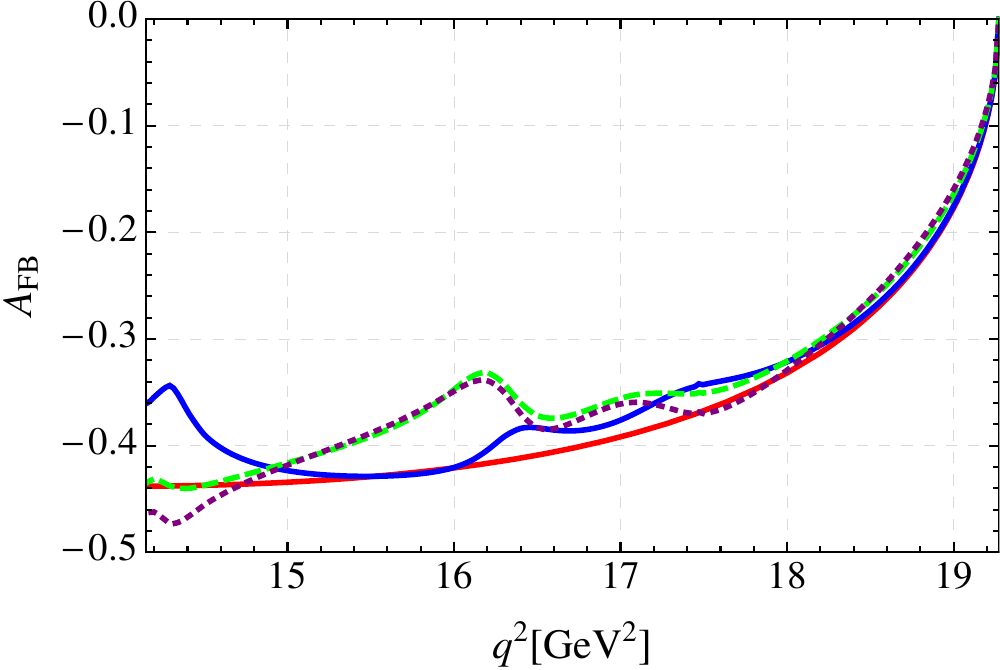}}
\subfigure{\includegraphics[width=0.380\textwidth]{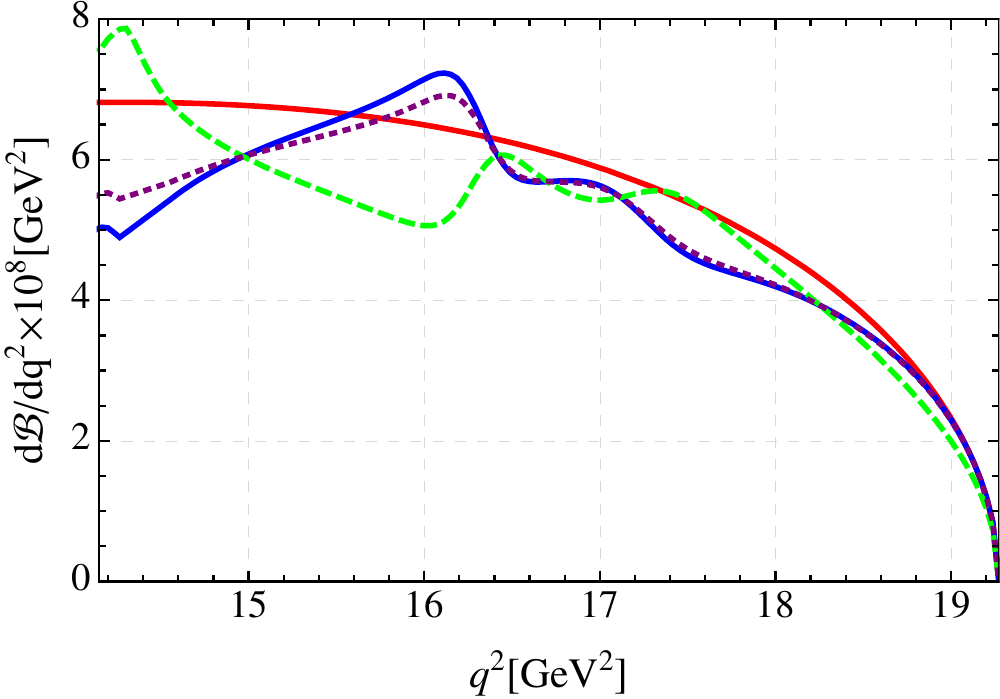}}
\caption{Local SM shapes of $A_{\rm FB}$ and $d{\cal{B}}/dq^2$  for $\eta_{0,\parallel,\perp}=1$  (solid blue curve) and $\eta_{0,\parallel,\perp}=-1$  (dashed green  curve), as well as with non-universality, $\eta_{\parallel,0}=1$, $\eta_\perp=-1$ (dotted purple curve).
The red curve without wiggles illustrates the unbinned OPE. To avoid clutter only theory curves using central values of input are shown.}
\label{S5contFA}
\end{center}
\end{figure}

The sensitivity of $A_{\rm FB}$ and the branching fraction to the resonance parameters is illustrated  in Fig.~\ref{S5contFA}.
($B \to K^* \ell \ell$ $q^2$-spectra including resonance effects have been given previously in ~\cite{Ali:1991is, Ali:1999mm, Ligeti:1995yz}, and
recently  in~\cite{Lyon:2014hpa} using $B\to K\mu\mu$ data.)
The sensitivity of  $S_5$ is very similar to the one of $A_{\rm FB}$  and not shown.
We recall that in these observables  already universal resonance effects do not cancel.
Shown are local SM spectra for  universal and constant $\eta_{0,\parallel,\perp}=\pm1$. This choice is consistent with the measured $B \to J/\Psi K^*$ and $B \to \Psi(2S) K^*$ branching ratios, see Appendix~\ref{sec:charm}.
In addition, we show the impact of non-universality, $\eta_{\parallel,0}=1$, $\eta_\perp=-1$ (dotted purple curve).
The resulting spread for different $\eta_j$ is rather small above $\sim 15 \,  \mbox{GeV}^2$ except in the branching ratio,
which could be used to detail the charm contribution locally, as in $B^+\to K^+\mu\mu$ decays~\cite{Aaij:2013pta}.
Ideally this should be done for each $K^*$ transversity state, 
$d \Gamma_j/d q^2$, see Eq.~(\ref{eq:br}). 

\section{Model-independent analysis}
\label{sec:SM}

\begin{figure}[H]
\begin{center}
\subfigure{\includegraphics[width=0.30\textwidth]{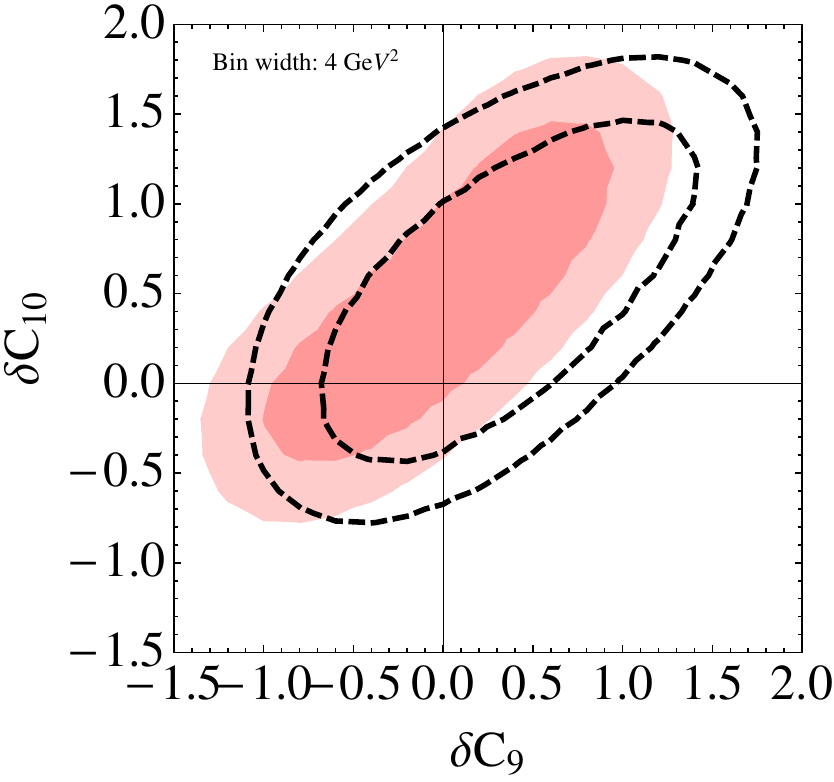}}
\subfigure{\includegraphics[width=0.30\textwidth]{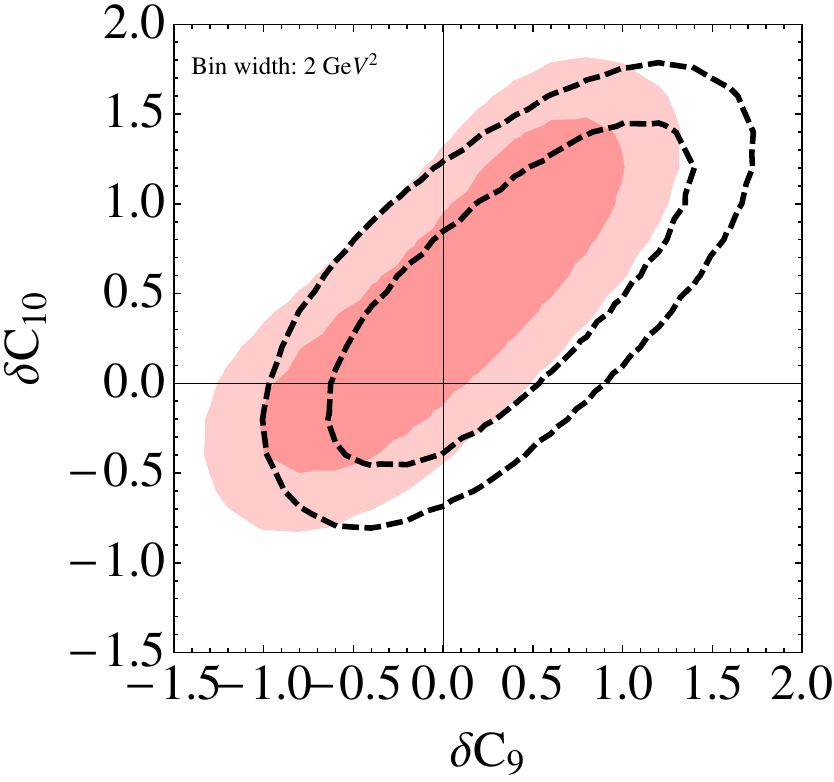}}
\subfigure{\includegraphics[width=0.30\textwidth]{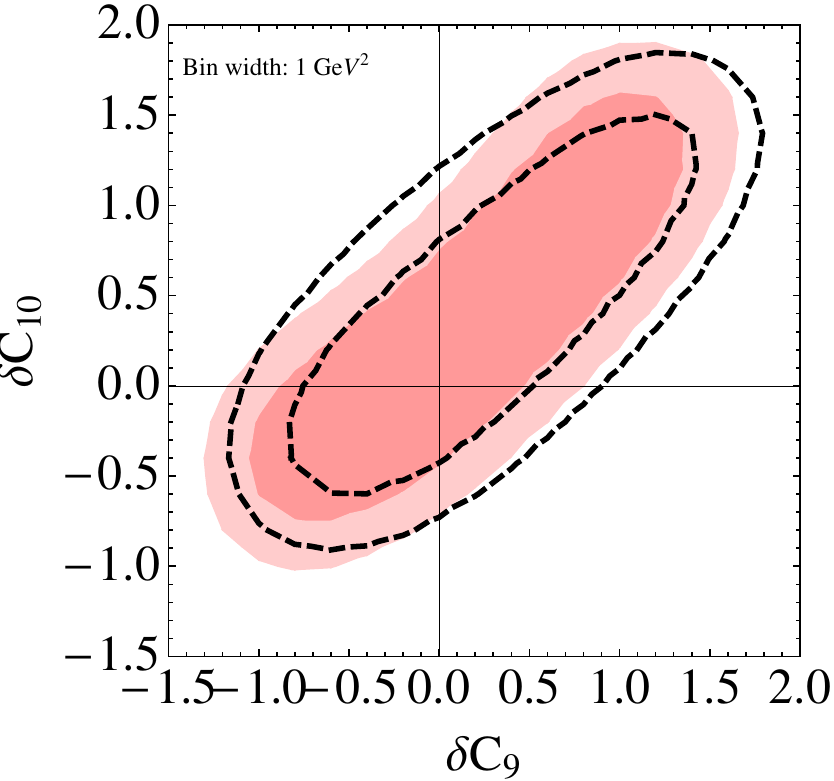}}
\subfigure{\includegraphics[width=0.30\textwidth]{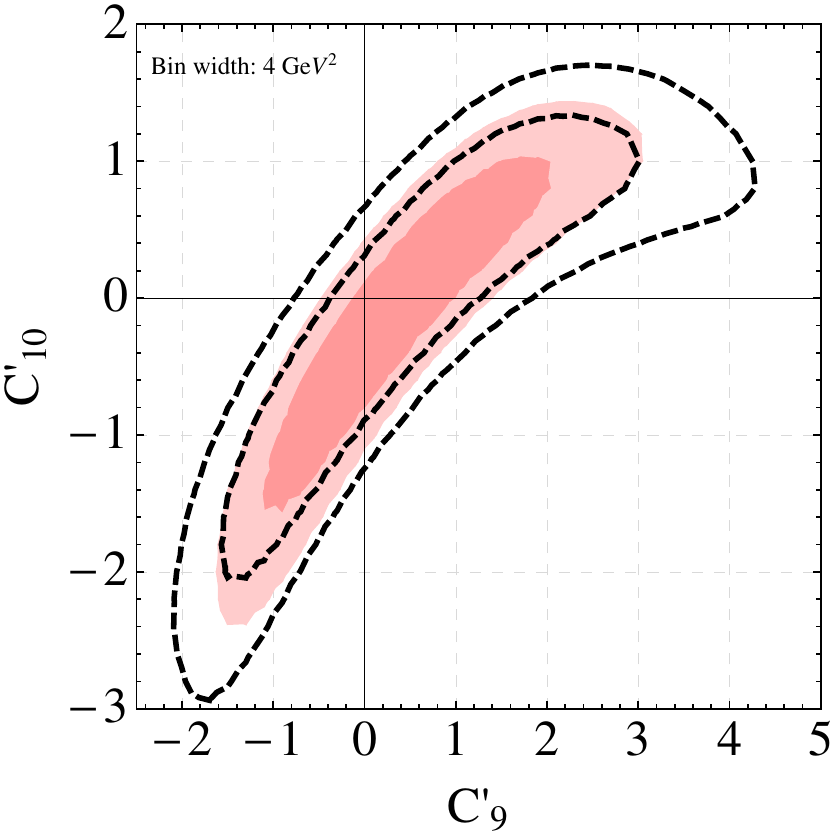}}
\subfigure{\includegraphics[width=0.30\textwidth]{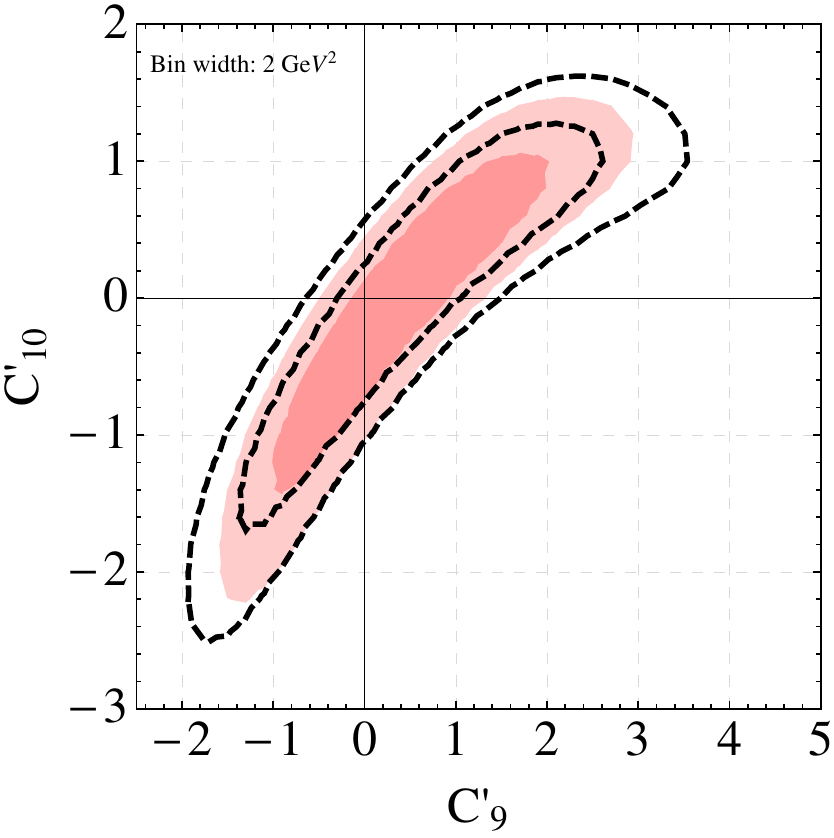}}
\subfigure{\includegraphics[width=0.30\textwidth]{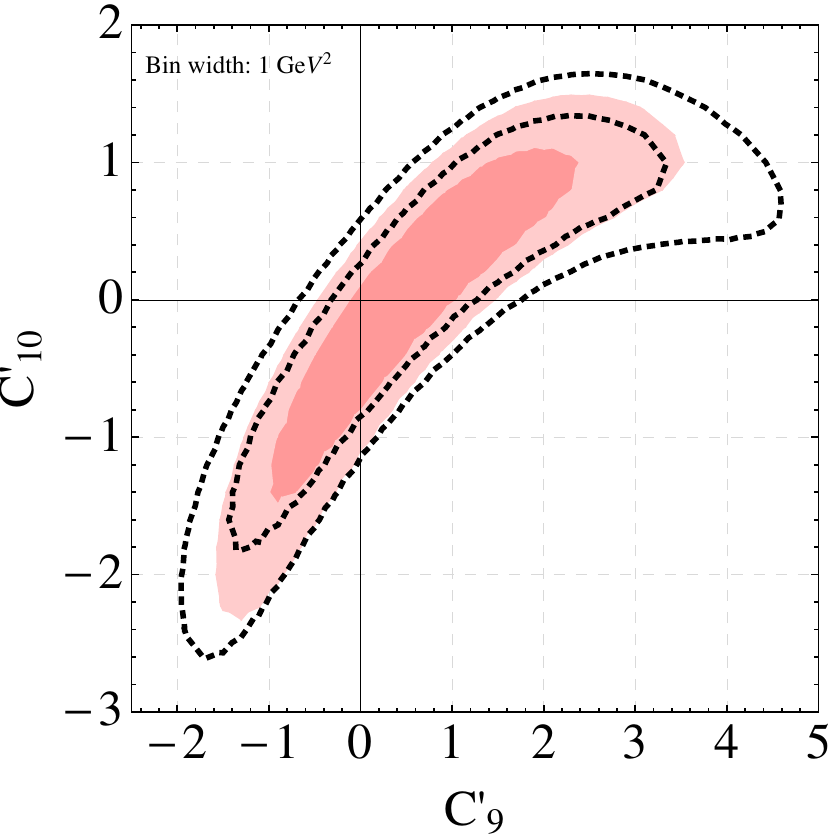}}
\caption{1 and 2 $\sigma$ constraints on the BSM coefficients  $ \delta \mathcal{C}_9$, $\delta \mathcal{C}_{10}$ (upper plots) and
 $ \mathcal{C}_9^\prime$, $ \mathcal{C}_{10}^\prime$ (lower   plots) from $B \to K^* \mu \mu$ decays at low recoil for different binning 
 4 $\mbox{GeV}^{2}$ (left), 2 $\mbox{GeV}^{2}$ (center) and 1 $\mbox{GeV}^{2}$ (right) as in 
 Eq.~(\ref{eq:bins}). Red shaded areas (dashed black contours) denote the allowed regions in the OPE (in the KS-approach with $\eta_\perp, \eta_0=\eta_\parallel$ simultaneously fitted),
 see text for details.}
\label{fig:Cfit}
\end{center}
\end{figure}

The constraints on the BSM coefficients from the different  fits using the three available $q^2$-binnings are presented in Fig.~\ref{fig:Cfit}.
We recall that all  fits are  based on $B \to K^* \mu \mu$ data at low recoil only.
The black dashed contours are from a simultaneous fit to Wilson coefficients and resonance parameters $\eta_0, \eta_\perp$ as discussed in Section~\ref{sec:uni}.
The red shaded areas are obtained within the OPE. In these plots, only two BSM coefficients are switched  on at a time, that is,
$ \delta \mathcal{C}_9$, $\delta \mathcal{C}_{10}$ in the upper plots and
 $ \mathcal{C}_9^\prime$, $ \mathcal{C}_{10}^\prime$ in the  lower   plots (Formulae which include right-handed currents can be taken from~\cite{Bobeth:2012vn}.).

We find that  within the OPE, as well as the local charm models,  the SM agrees well with the data at the current level of precision. 
The findings are  consistent with the pure low recoil analysis of Ref.~\cite{Descotes-Genon:2015uva}.
Zooming in from large to small bins, the OPE result undergoes small changes, caused by the binning-dependent  experimental uncertainties.
With the local KS-model zooming in increases the resolution to charmonium contributions. In the fits for $C_{9,10}$ we find $\chi^2/d.o.f.=(1.3, 0.8, 1.3)$ within the OPE and $(1.0, 0.6, 1.2)$ within the KS-model (simultaneously fitting for $C_{9,10}$ and $\eta_{0,\perp}$)  for $(1,2,4)\,\text{GeV}^2$ bins, respectively. Similar results are obtained for the corresponding fits to $C'_{9,10}$.
All plots exhibit consistency between local modelling and the OPE. We conclude that
within current precision, charm effects appear to be controlled and do not endanger the validity of BSM constrains.

To estimate the uncertainties of the OPE predictions for a given binning, we suggest to use the ratios
\begin{align} \label{eq:ek}
\epsilon_1 = \frac{ \int_{bin} \rho_1^{KS}  dq^2}{ \int_{bin} \rho_1^{OPE}  dq^2} \, ,  \quad 
\epsilon_2 = \frac{ \int_{bin} \rho_2^{KS}  dq^2}{ \int_{bin} \rho_2^{OPE}  dq^2} \, , \quad  ~~
\epsilon_{12} = \frac{ \int_{bin} \rho_2^{KS}  dq^2}{ \int_{bin} \rho_2^{OPE}  dq^2} \cdot    \frac{ \int_{bin} \rho_1^{OPE}  dq^2}{ \int_{bin} \rho_1^{KS}  dq^2} \, ,
\end{align}
where   $\rho_{1,2}$ are given in Eqs.~(\ref{eq:ro1}) and (\ref{eq:ro2})
and evaluated with the respective $\mathcal{C}_9^{\rm eff}$. The $\epsilon_k$ are theory measures of the OPE's binning-related uncertainty.
Their relation to the observables is straight-forward in the universality-limit.

The $\epsilon_k$ are worked out in Table~\ref{tab:error}, taking into account the 
$1 \sigma$ ranges of $\eta_{0,\perp}, C_{9,10}$ of the fit shown in Fig.~\ref{fig:NU-fit}.
\begin{table}[H]
\begin{center}
\begin{tabular}{c||c||c|c||c|c|c|c}
bin in $\mbox{GeV}^{2}$ &  $15-19$ & $15-17 $ & $17-19$ &  $15-16$ & $16-17$& $17-18$ & $18-19$  \\ \hline
$\epsilon_1$  &(0.85,1.16) & (0.81,1.30&(0.87,1.03) &(0.76,1.20) &(0.84,1.38) &(0.84,1.03) &(0.86,1.05)\\
$\epsilon_2$  &(0.82,1.0) &(0.74,1.13) &(0.85,0.91) &(0.71,1.17) & (0.78,1.08)&(0.76,0.95) &(0.84,0.97)\\
$\epsilon_{12}$  &(0.86,1.05) &(0.87,1.05) & (0.84,1.05)&(0.95,1.06) &(0.78,1.05) &(0.75,1.05) & (0.93,1.05)\\
\end{tabular}
 \caption{Ratios $\epsilon_k$ defined in Eq.~(\ref{eq:ek}) for different $q^2$-bins and $1 \sigma$ ranges of  parameters $\eta_0,\eta_\perp$ and  $\mathcal{C}_{9,10}$. The coefficients $\mathcal{C}'_{9,10}$ are set to zero.}
  \label{tab:error}
 \end{center}
\end{table}  
As expected, larger bins are better behaved than smaller ones, {\it i.e.,}\@~have $\epsilon_k$ closer to 1, except for those near the endpoint.
The $[17-19]\,\mbox{GeV}^2$ and the  $[18-19]\, \mbox{GeV}^2$ one  are preferable to the $[15-19]\,\mbox{GeV}^2$ bin.
We also learn that the corrections to $\rho_2/\rho_1$, $\epsilon_{12}$, are not always  favored with respect to  $\epsilon_1$ or $\epsilon_2$, caused by inefficient
cancellation of charm effects. 
Presently deviations of  around 30 \% exist in  $[15-16]\,\mbox{GeV}^2$, $[16-17]\,\mbox{GeV}^2$ and the  $[15-17]\,\mbox{GeV}^2$ bins.
The deviations for the preferred  bins are at most  16 \%, and directed towards reducing KS- versus OPE-distributions.
In Table~\ref{tab:error} mostly   $\epsilon_k <1$.
This is driven by ${\rm Re }[h_c(q^2)] < {\rm Re} [h(q^2,m_c^2)]$ above $q^2 \sim 16.4\,\mbox{GeV}^{2}$, see Fig.~\ref{Fig.1}.

If the resonance parameters $\eta_c$ would be determined more precisely,
the uncertainty on the mismatch between the OPE and the KS-model would shrink.
This way, the deviations $|\epsilon_k-1|$ given here correspond to upper limits.

Note the possible ambiguity that could arise if there is a significant constant or slowly varying contribution to $\mathcal{C}^{\text{eff}}_9$ stemming from  non-resonant 
$D\bar{D}$-backgrounds. Such effect might be differentiated from a new physics contribution to  $\mathcal{C}_9$ only in the case the latter is CP-violating or lepton flavor non-universal. Hadronic backgrounds not captured by the OPE for a given binning cause the KS-fit and the OPE-fit to disagree.
We interpret this as an uncertainty  of the OPE-fit to the Wilson coefficients. A possible binning-independent uncertainty will be one of the limiting factors to test the SM at low recoil.

\section{Conclusions}
\label{sec:con}

The low recoil region in semileptonic  $|\Delta B| =|\Delta S|=1$ decays is inhabited by wider charm resonances,  locally not captured by the OPE.
While with current data such resonance patterns may be only at the border of being visible in $B \to K^* \mu \mu$ decays, see Figs.~\ref{PlotsFLcont} and  \ref{fig:S5AFB},
in the near future this will be an important  background to SM precision tests.
In order to understand these effects,  which are inaccessible from within the OPE, 
we use the KS-model \cite{Kruger:1996cv},  cf. Section \ref{sec:KS}, which does describe resonances locally, as a test-case  against the OPE .

Using available $B \to K^* \mu \mu$ data at low recoil, we  performed simultaneous fits to resonance parameters  of the KS-model and BSM Wilson coefficients, and compare it to the plain OPE-fit.
We find that the resulting constraints are consistent with each other, and  consistent with the SM, see  Fig.~\ref{fig:Cfit}.
There is room  left for sizeable BSM contributions. Let us emphasize the difference between our work and the recent global fits~\cite{Descotes-Genon:2015uva, Hurth:2014vma, Beaujean:2013soa} that used the experimental data for all $b\to s$ processes as an input but included only the total low recoil bin for $B\to K^\ast\ell\ell$. We focus solely on the low recoil region of this decay mode and use all available data for this kinematic region including also the smaller bins, since our goal is to scrutinize the local $q^2$-shape.
Specifically, in the KS-model fit we do not fix the hadronic parameters expressed by the fudge factors but use them as  fit parameters together with the short-distance Wilson coefficients.

To estimate the uncertainties of the OPE for a given binning, we use the coefficients $\epsilon_k$ defined in Eq.~(\ref{eq:ek}), with 
current evaluations shown in Table \ref{tab:error}.  Preferred 
are the endpoint bins,  $[17-19] \,  \mbox{GeV}^2$ and $[18-19] \,  \mbox{GeV}^2$ followed by the large one $[15-19] \, \mbox{GeV}^2$. 
For the $[17-19] \,  \mbox{GeV}^2$ bin, we find model-independently (strongly directional) deviations from  the OPE not exceeding 15 \%.

In the future more precise data  with even smaller binning than currently exist would be desirable 
 to determine the resonance parameters more accurately. This will directly influence
the estimates in Table \ref{tab:error},  which  measure  not only the mismatch between local spectra and the OPE, but include also uncertainties within the KS-model.
Refinements of the method, such as less minimal parameterizations for the $\eta_c(K^*_j,q^2)$ and $h_c(q^2)$ functions, may also be envisaged.
As dominant uncertainties within the OPE are due to hadronic form factors,
improving their predictions would be desirable, too.

\section*{Acknowledgements}
Authors would like to thank Damir Be\v cirevi\' c  and the participants of the MITP-workshop "Flavour and Electroweak Symmetry Breaking" at Capri island, June 13-24, 2016  for useful discussions. This project is supported in part 
by the {Bundesministerium f\"ur Bildung und Forschung (BMBF)}.

\begin{appendix}
\section{The angular coefficients \label{sec:Ji}}
The angular coefficients in Eq.\eqref{Angular distributions} are given in terms of the transversity amplitudes as follows
\begin{equation}
\small
 \begin{split}
 & J_1^s =\frac{9}{16}\bigg[\vert\mathcal{A}_\perp^L\vert^2+\vert\mathcal{A}_\parallel^L\vert^2+(L\rightarrow R)\bigg] \quad J_1^c =\frac{3}{4}\bigg[\vert\mathcal{A}_0^L\vert^2+(L\rightarrow R)\bigg] \quad J_2^s =\frac{3}{16}\bigg[\vert\mathcal{A}_\perp^L\vert^2+\vert\mathcal{A}_\parallel^L\vert^2+(L\rightarrow R)\bigg] \\
 & J_2^c =-\frac{3}{4}\bigg[\vert\mathcal{A}_0^L\vert^2+(L\rightarrow R)\bigg]\quad J_3  =\frac{3}{8}\bigg[\vert\mathcal{A}_\perp^L\vert^2-\vert\mathcal{A}_\parallel^L\vert^2+(L\rightarrow R)\bigg]\quad
 J_4 = \frac{3}{4\sqrt{2}}\bigg[\operatorname{Re}(\mathcal{A}_0^L\mathcal{A}_\parallel^{L\,\ast})+(L\rightarrow R) \bigg]\\
 & J_5 =\frac{3\sqrt{2}}{4}\bigg[\operatorname{Re}(\mathcal{A}_0^L\mathcal{A}_\perp^{L\,\ast})-(L\rightarrow R) \bigg]\quad J_6 =\frac{3}{2}\bigg[\operatorname{Re}(\mathcal{A}_\parallel^L\mathcal{A}_\perp^{L\,\ast})-(L\rightarrow R) \bigg]\quad
J_7 =\frac{3\sqrt{2}}{4}\bigg[\operatorname{Im}(\mathcal{A}_0^L\mathcal{A}_\parallel^{L\,\ast})-(L\rightarrow R) \bigg]\\
& J_8 =\frac{3}{4\sqrt{2}}\bigg[\operatorname{Im}(\mathcal{A}_0^L\mathcal{A}_\perp^{L\,\ast})+(L\rightarrow R) \bigg]\quad
J_9 =\frac{3}{4}\bigg[\operatorname{Im}(\mathcal{A}_\parallel^L\mathcal{A}_\perp^{L\,\ast})+(L\rightarrow R) \bigg],\\  
 \end{split}
\end{equation}
where we neglected terms proportional to $m_\ell^2/q^2$. The full expressions can be found in Ref.\cite{Bobeth:2010wg}.

\section{Fitting $R(q^2)$}\label{Details of the fit for $R(q^2)$}

Here we describe the fitting procedure for the ratio $R(q^2)$ defined in Eq.~\eqref{Rfit} using the experimental input on the $e^+e^-\to h_i$ cross section from the BES-II experiment~\cite{Ablikim:2007gd}. Above the $\bar{D}D$ threshold the four wide charmonium resonances, $\psi(3770), \psi(4040), \psi(4150)$ and $\psi(4415)$, with quantum numbers $J^{PC}=1^{--}$, appear in the spectrum. We adopt the fitting procedure from~\cite{Ablikim:2007gd} and model the background according to~\cite{Lyon:2014hpa}. 

\begin{figure}[H]
\label{RPlot}
\begin{center}
\subfigure{\includegraphics[width=0.49\textwidth]{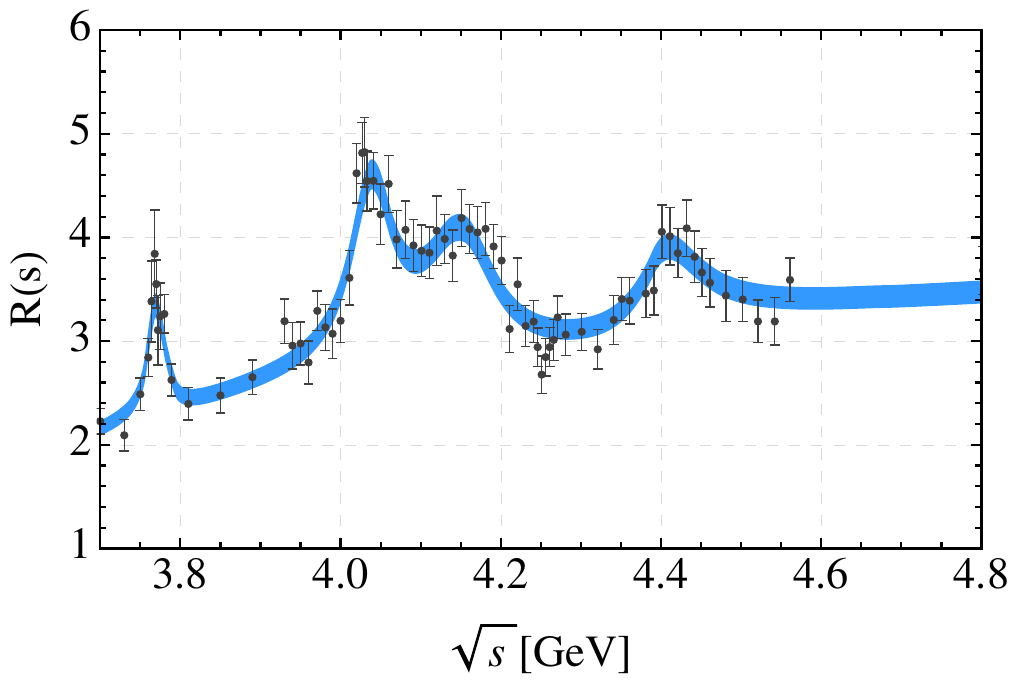}}
\caption{The result of the fit for the function $R(q^2)$ (blue  1  $ \sigma$ band) defined  in Eq.~\eqref{Rfit} using the experimental input on the $e^+e^-\to h_i$ cross section from the BES-II experiment \cite{Ablikim:2007gd}.} \label{fig:Rfit}
\end{center}
\end{figure}

The transition amplitude of the resonance $r$ into a final state $f$ is modelled by the Breit-Wigner ansatz with a phase $\delta_r$ and mass $m_r$:
\begin{equation}
\mathcal{T}^{r\to f}=\frac{m_r\sqrt{\Gamma^{r\rightarrow e^+ e^-} \Gamma^{r\rightarrow f}(s)}}{s-m_r^2+i\,m_r\Gamma_r(s)}e^{i\delta_r}.
\end{equation}  
Since only three relative phases carry a physical information, we set $\delta_{\psi(3770)}=0$. The $s$-dependent decay widths   of $r \to f$ 
are given by the formula
\begin{equation}
\Gamma^{r\rightarrow f}(s)=\bar{\Gamma}_r\frac{2m_r}{m_r+\sqrt{s}}\sum_L \frac{Z_f^{2L+1}}{B_L} \, .
\end{equation} 
Here, $\bar{\Gamma}_r$ denotes a fit parameter specified for every given resonance and $Z_f\equiv \rho P_f$, where$\rho\simeq\,1\mbox{GeV}$. The sum over the orbital angular momenta of the decaying final states and the energy dependent partial wave function $B_L$ are given in~\cite{Ablikim:2007gd}. The momentum $P_f$ of the two body decay of the resonance $f$ into the final mesons with masses $m_1$ and $m_2$ is given by the familiar formula
\begin{equation}
P_f=\frac{\sqrt{\lambda(m_1^2,m_2^2,m_r^2)}}{2m_r} \, . 
\end{equation}
The different decay channels are given in Table \ref{tab:ResTab}. The total hadronic width is the sum over all final states
\begin{equation}
\Gamma_r^{\text{had}}(s)=\sum_f\Gamma^{r\rightarrow f}(s) \, .
\end{equation}
The total width of a charmonium resonance is the sum of the hadronic and leptonic widths
\begin{equation}
\Gamma_r(s)=\Gamma^{r\rightarrow e^+ e^-}+\Gamma^{r\rightarrow \mu^+ \mu^-}+\Gamma^{r\rightarrow \tau^+ \tau^-}+\Gamma_r^{\text{had}} \, .
\end{equation}
Lepton universality $\Gamma^{r\rightarrow e^+ e^-}=\Gamma^{r\rightarrow \mu^+ \mu^-}$ is assumed for the leptonic decay widths of the electron and the muon. The kinematic suppression factors are included for the decay rates of the resonances that involve tau pairs in the final state. The total square of the modulus of the inclusive amplitude of the resonances is the following incoherent sum over the final states:
\begin{equation}
\vert\mathcal{T}\vert^2=\sum_f\bigg\vert \sum_r \mathcal{T}^{r\rightarrow f}(s)\bigg\vert^2 \, .
\end{equation}
The resonance contribution $R_{\text{res}}(s)$ to $R(s)$ (see Eq.~\eqref{Rfit2}) is given by
\begin{equation}
R_{\text{res}}(s)=\frac{9}{\alpha^2_{\text{em}}}  \vert\mathcal{T}_{\text{res}}\vert^2 \, .
\end{equation}
The continuum background is modelled as~\cite{Lyon:2014hpa}
\begin{equation}
R_{\text{cont}}(s)=R_{uds}+\theta(s- 4 m_D^2)(1-x) (\Delta R_c+ x a_{\text{cont}}),\quad \Delta R_c=R_{udsc}-R_{uds},
\end{equation}
with $x=4m_D^2/s$. For the light quark ratio $R_{uds}$ and the one including charm, $R_{udsc}$, we use the predictions
from Refs.~\cite{Harlander:2002ur,Kuhn:2007tc}. Specifically, we employ
$R_{uds}=R(s=(3.73\,\text{GeV})^2)$ and $R_{udsc}=R(s=(4.8\,\text{GeV})^2)$. 
The result of the fit for $R(s)$ in  the fit interval $\sqrt{s}=(3.7, 4.8)\,\text{GeV}$ is shown in Fig.~\ref{fig:Rfit}. We use $76$ data points for $R(s)$ and find $\chi^2/d.o.f=1.01$, for $d.o.f.=76-17-1$. Our fit results are consistent with the Ref.~\cite{Ablikim:2007gd}.
The charm contribution $R_c(s)$ can be extracted from Eq.~\eqref{Rc}.

To evaluate  the dispersion integral \eqref{dispersion relation} also below the fit interval, we require  contributions to $R(s)$ from the narrow resonances $J/\psi$ and $\psi(2S)$, parameterized  as
\begin{equation}
h_{c,\text{narrow}}(s)=-\frac{3\pi}{\alpha_{\text{em}}^2} \sum_{r = J/\psi,\psi(2S)}\frac{m_r \Gamma^{r\rightarrow\ell^+\ell^-}}{s-m_r^2+im_r\Gamma_r} \,  . \label{below}
\end{equation} 
\begin{table}
 \begin{center}
\begin{tabular}{p{1.8cm}p{1cm}p{1cm}p{1cm}p{1cm}p{1cm}p{1cm}p{1cm}p{1cm}}
    {$\psi(3770)\rightarrow$} & {$D\bar{D}$} & {} & {} & {} & {} & {} &{} & {}  \\ \midrule
    \hline
    {$\psi(4040) \rightarrow$} &{$D\bar{D}$} & {$D^\ast \bar{D}^\ast$} & {$D\bar{D}^\ast$} & {$D_s\bar{D}_s$} & {} & {} & {} & {}  \\ \midrule
    \hline
  {$\psi(4140)\rightarrow$} & {$D\bar{D}$} & {$D^\ast \bar{D}^\ast$} & {$D \bar{D}^\ast$} & {$D_s \bar{D}_s$} & {$D_s\bar{D}^\ast_s$} & {} & {} & {}  \\ \midrule
  \hline
  {$\psi(4415)\rightarrow$} & {$D\bar{D}$} & {$D^\ast \bar{D}^\ast$} & {$D \bar{D}^\ast$} & {$D_s \bar{D}_s$} & {$D_s \bar{D}_s^\ast$} & {$D_s^\ast \bar{D}_s^\ast$} &{$D \bar{D}_1$} & {$D \bar{D}_2^\ast$}\\\bottomrule 
\end{tabular}
 \caption{Two body decays of charm resonances into final states $\emph{f}$. \label{tab:ResTab}}
 \end{center}
\end{table}  
The values of the parameters in \eqref{below} used in the fit are given in the Table \ref{TabPDG}. 
Above the fit interval in the open charm region, we use the Schwinger's $\mathcal{O}(\alpha_s)$ result, in the form adopted from~\cite{Lyon:2014hpa},
\begin{equation}
\operatorname{Im}\left[ h_{c, \text{above}}(s)\right]=\frac{2\pi}{9}(3-\beta^2(s))\vert \beta(s)\vert\bigg[1+\frac{4}{3}\alpha_s\bigg(\frac{\pi}{2\beta(s)}-\bigg(\frac{3}{4}+\frac{\beta(s)}{4}\bigg)\big(\frac{\pi}{12}-\frac{3}{4\pi})\bigg)\bigg],
\end{equation}
where $\beta(s)=\sqrt{1-4m_c^2/s}$. 
The final result for  $h_c(q^2)$ is given in Fig.~\ref{Fig.1}.

\begin{table}[H]
\begin{center}
\begin{tabular}{p{1.cm}p{2.6cm}p{2.0cm}p{2.2cm}}
    {$r$} & {$m_r/\text{MeV}$} & {$\Gamma_r/\text{keV}$} & {$\Gamma^{r\rightarrow e^+e^-}/\text{keV}$} \\ \hline
    {$J/\psi$} &{$3096.916 \pm 0.011$} & {$92.9\pm 2.8$} & {$5.55\pm 0.14$} \\ \hline
  {$\psi(2S)$} &{$3686.109 \pm 0.013$} & {$299\pm 8$} & {$2.36\pm 0.04$} \\ \hline
\end{tabular}
 \caption{The values of the parameters for the narrow charm resonances taken from \cite{Agashe:2014kda} used in the fit.}\label{TabPDG}
 \end{center}
\end{table}  

\section{ $B \to \psi_i K^{(*)}$ data and   factorization\label{sec:charm}}

Here we give a brief overview on phenomenological data on  $B \to \psi_i K^{(*)}$ decays, where $\psi_i$ denotes a generic charmonium  $1^{--}$ resonance, within factorization.
The matrix element of the $B \to \psi_i K^{(*)}$ decays can be written as
\begin{equation}
\mathcal{M}(B\to \psi_i K^{(*)})=\frac{4 G_F}{\sqrt{2}}V^{\ast}_{cb}V_{cs}  \langle \psi_i K^{(*)} \vert \mathcal{C}^{(1)} \mathcal{O}^{(1)} + \mathcal{C}^{(8)} \mathcal{O}^{(8)}  \vert B\rangle \, ,
\end{equation}
where the  commonly used color singlet and octet operators read
\begin{equation}
\mathcal{O}^{(1)}=(\bar{c}\gamma_\mu P_Lc)(\bar{s}\gamma^\mu P_L b),\,\, \mathcal{O}^{(8)}=(\bar{c}T^a\gamma_\mu P_L c)(\bar{s}T^a\gamma^\mu P_L b)   \, , \label{Ops}
\end{equation}
respectively.
The Wilson coefficients of these operators read  in terms of the ones  in the  CCM-basis (Eq.~(\ref{EffHamiltonian1})) as
$\mathcal{C}^{(1)}=  a_2$, given in  Eq.~(\ref{eq:a2}),  and $\mathcal{C}^{(8)}=\frac{1}{3}\left(-\mathcal{C}_1+6\mathcal{C}_2\right)$.
We employ the value of  $a_2$ at NNLO~\cite{Gorbahn:2004my, Bobeth:1999mk} thereby including perturbative corrections to the weak $b \to c \bar c s$ vertex.

Assuming factorization, the $B \to \psi_i K^{(*)}$ matrix element reads 
\begin{equation}
\mathcal{M}_{\textbf{fac}}(B\to \psi_i K^{(*)})=\frac{G_F}{\sqrt{2}}V_{cb}^\ast V_{cs} a_2 \kappa \langle \psi_i \vert  \bar{c}\gamma_\mu c \vert 0\rangle\langle K^{(*)}\vert\bar{s}\gamma^\mu(1-\gamma_5)b\vert B\rangle.\label{afacK}
\end{equation}
For  $\kappa=1$ this  \emph{ansatz}   represents the naive factorization approximation (NFA).
Note that the  color octet operator does not contribute, {\it e.g.}, ~\cite{Boos:2004xp}.
The dependence on the renormalization scale $\mu$ does not cancel between the Wilson coefficients and the effective operators in the \emph{ansatz} \eqref{afacK}.
For further aspects of the factorization ansatz the reader is referred to \cite{Neubert:1997uc}.
The matrix element in factorization can be expressed in terms of charmonium decays constants,  which can be extracted from  data on $\Gamma(\psi_i\to \ell\ell)$ and
form factors, to be evaluated at $q^2=m_{\psi^2_i}$ 
\begin{equation}
\begin{split}
 \langle \psi_i (q, \epsilon)  \vert \bar{c}\gamma_\mu c\vert 0\rangle&=i\,f_{\psi_i}m_{\psi_i}\epsilon_{\mu}^\ast,\\
  \langle K(k) \vert \bar{s}\gamma_\mu b\vert B(p)\rangle&=f_+(q^2)\bigg( (p +k)_{\mu}-\frac{m_B^2-m_K^2}{q^2}q_\mu\bigg) +f_0 (q^2) \frac{m_B^2-m_K^2}{q^2}q_\mu\ \, , 
\end{split}
\end{equation}
\begin{align}
    \langle K^*(k, \eta)  \vert  \bar{s} \gamma_\mu b  \vert  B(p)   \rangle 
        & = \frac{2 V(q^2)}{m_B + m_{K^*}} \varepsilon_{\mu\rho\sigma\tau} \eta^{*\rho} p^\sigma k^\tau ,\\
    \langle K^*(k, \eta) \vert  \bar{s} \gamma_\mu \gamma_5 b \vert  B(p) \rangle  
           & = i \eta^{*\rho}  \left[ 2 m_{K^*} A_0(q^2) \frac{q_\mu q_\rho}{q^2} + (m_B + m_{K^*}) A_1(q^2) \left( g_{\mu\rho} - \frac{q_\mu q_\rho}{q^2} \right) \right.
          \nonumber \\
       & - \left.A_2(q^2) \frac{q_\rho}{m_B + m_{K^*}}  \left(  (p+k)_\mu - \frac{m_B^2 - m_{K^*}^2}{q^2} q_\mu \right)   \right]  \, , 
\end{align}
where $\eta$ ($\epsilon$) denotes the $K^*$ ($\psi_i$) polarization vector,  $k,p$ the 4-momenta of the
$K^{(*)}$,  $\bar B$ mesons, respectively, and $q=p-k$.  Due to $\epsilon \cdot q=0$ the terms proportional to $q_\mu$ do not contribute.
We use $f_{J/\psi}=0.416 \pm 0.006$ GeV, $f_{\psi(2S)}=0.297 \pm 0.003$ GeV  and 
$f_{\psi(3770)}=0.100 \pm 0.004$ GeV \cite{Amhis:2014hma}.

There are several processes for which the measurements reveal deviations from  NFA. For instance, the branching fraction of the process $B^{-}\to K^{-}\chi_{c0}$ has been observed to be significantly non-vanishing, while the corresponding factorization contribution vanishes due to parity conservation of QCD. As a second example, the branching fraction of  $B\to K J/\psi$ also deviates from its NFA value,  to be discussed in Section \ref{sec:Kpi}.
A possible source of  non-factorizable corrections are $B$-meson decays to $D_s^{(\ast)}D^{(\ast)}$ pairs which afterwards re-scatter into  $K\psi_i$ pairs. The analysis of such effects was undertaken in Ref. \cite{Colangelo:2003sa}. It contains significant theoretical uncertainty related to the vague knowledge of the relevant strongly coupled meson vertices. In the following sections we follow the phenomenological point of view and extract the fudge factors for the processes $B\to K^{(\ast)}(J/\psi, \psi(2S))$, in order to gain some further insights.

We stress that for form factors from lattice  QCD  in the intermediate $q^2$-region around peaking charmonium resonances additional uncertainties apply.

\subsection{$B\to K (J/\psi,\psi(2S))$ \label{sec:Kpi}}

The $B\to\psi_i K$ branching ratio can be written as
\begin{equation}
\mathcal{B}(B\to \psi_i K)=\tau_{B}\frac{G_F^2\vert V_{cb}^\ast V_{cs}\vert^2}{32\pi m_B^3}a_2^2\, |\kappa_{\psi_i \,K}|^2 f_{\psi_i}^2\lambda^{3/2}(m_B^2,m_{\psi_i}^2,m_K^2)[f_+(m_{\psi_i}^2)]^2.
\end{equation}
 Using \cite{Agashe:2014kda}
\begin{align}
\mathcal{B}(\bar{B}^0\to J/\psi K^0)&=(0.873\pm 0.032)\times 10^{-3},\quad \mathcal{B}({B}^-\to J/\psi K^-)=(1.026\pm 0.031)\times 10^{-3} \, , \\
\mathcal{B}(\bar{B}^0\to \psi(2S) K^0)&=(0.58\pm 0.05)\times 10^{-3},\quad \mathcal{B}({B}^-\to \psi(2S) K^-)=(0.626\pm 0.024)\times 10^{-3}, \\
& \mathcal{B}({B}^-\to \psi(3770) K^-)=(0.49\pm 0.13)\times 10^{-3}, 
\end{align}
we obtain, after error-weighted averaging of neutral and charged $B$ decay modes if applicable, the following coefficients
\begin{equation}
\vert\kappa_{J/\psi\,K}\vert=1.40\pm 0.09,\quad \vert\kappa_{\psi(2S)\,K}\vert=1.72\pm 0.08 \, ,  \quad \vert\kappa_{\psi(3770)\,K}\vert=4.54 \pm 0.68 \, .  \label{kappaK} 
\end{equation}
Here, we used the form factor $f_+(m_{\psi_i}^2)$ evaluated in Lattice QCD in \cite{Bouchard:2013pna}, see also~\cite{Bailey:2015dka}.
Form factors extrapolated to $q^2= m_{\psi_i}^2$ from light cone sum rules  from \cite{Khodjamirian:2010vf} yield very similar results.
The value for $\psi(3770)$ is only given for completeness as this resonance is included in the fit to BES-data.
The values in  Eq.~(\ref{kappaK})  reveal an order one deviation from naive factorization, $\kappa=1$. Note, that the corresponding signs (phases) remain undetermined from this extraction.

\subsection{$B\to K^\ast(J/\psi,\psi(2S))$  \label{sec:Kst}}

Assuming universal  fudge factors for all polarizations of the $K^\ast$, where $ \kappa_{\psi_i\,K^\ast} =\eta_c(K^*, m_{\psi_i}^2)$, the $B\to\psi_i K^*$ branching fraction can be written as
\begin{equation}
\begin{split}
\mathcal{B}(B\to \psi_i K^\ast)&=\tau_B\frac{G_F^2\vert V_{cb}^\ast V_{cs}\vert^2}{32\pi m_B^3}a_2^2\, |\kappa_{\psi_i \,K^\ast}|^2 f_{\psi_i}^2\lambda^{1/2}(m_B^2,m_{\psi_i}^2,m_{K^\ast}^2)(m_B+m_{K^\ast})^2m_{\psi_i}^2[A_1(m_{\psi_i}^2)]^2 \\
&\times\bigg[(a-bx)^2+2(1+c^2 y^2)\bigg],
\end{split}
\end{equation}
where 
\begin{equation}
a=\frac{m_B^2-m_{K^\ast}^2-m_{\psi_i}^2}{2m_{K^\ast}m_{\psi_i}},\, \quad b=\frac{\lambda(m_B^2,m_{K^\ast}^2,m_{\psi_i}^2)}{2m_{K^\ast}m_{\psi_i}(m_B+m_{K^\ast})^2},\ \quad c=\frac{\lambda^{1/2}(m_B^2,m_{K^\ast}^2,m_{\psi_i}^2)}{(m_B+m_{K^\ast})^2},
\end{equation}
and the ratios of the $B\to K^\ast$ form factors
\begin{equation}
x=\frac{A_2(m_{\psi_i}^2)}{A_1(m_{\psi_i}^2)},\, \quad \quad  y=\frac{V(m_{\psi_i}^2)}{A_1(m_{\psi_i}^2)} \, .
\end{equation}
Using \cite{Agashe:2014kda}
\begin{align}
\mathcal{B}(\bar{B}^0\to J/\psi K^{0\ast})&=(1.32\pm 0.06)\times 10^{-3},\quad \mathcal{B}({B}^-\to J/\psi K^{-\ast})=(1.43\pm 0.08)\times 10^{-3} \, , \\
\mathcal{B}(\bar{B}^0\to \psi(2S) K^{0\ast})&=(0.59\pm 0.04)\times 10^{-3},\quad \mathcal{B}({B}^-\to \psi(2S) K^{-\ast})=(0.67\pm 0.14)\times 10^{-3}
\end{align}
we obtain after averaging over neutral and charged $B$ decay modes values of $|\kappa|$ close to unity:
\begin{equation}
\vert\kappa_{J/\psi\,K^\ast}\vert=0.96\pm 0.06,\quad \vert\kappa_{\psi(2S)\,K^\ast}\vert=0.85\pm 0.06.\quad\label{kappaKst}
\end{equation}
Here, we employed the $B\to K^\ast$ form factors from \cite{Horgan:2013hoa}.
Eqs.~\eqref{kappaK} and  \eqref{kappaKst}  suggest that  non-factorizable corrections for  processes involving a $K^\ast$ are smaller than those with a  $K$. 

Further information on universality can be obtained from data on the polarization fractions of the $K^*$ in  $B\to K^\ast(J/\psi,\psi(2S))$ decays  \cite{Aubert:2007hz,Aaij:2013cma,Hiller:2013cza}
\begin{align}
|A_\perp|^2= 2  \frac{c^2 y^2}{r} \, , \quad |A_\parallel |^2= \frac{2}{r} \, , \quad 
r\equiv  (a-bx)^2+2(1+c^2 y^2) \,  ,
\end{align}
which are normalized as $|A_0|^2+|A_\parallel|^2+|A_\perp|^2=1$. We compare data to the factorization predictions  in Table \ref{tab:Kstpol}
for form factors extrapolated  from the lattice  \cite{Horgan:2013hoa} and a phenomenological fit (SE2LEL)  \cite{Hambrock:2013zya}.
Results based on \cite{Khodjamirian:2010vf} have larger uncertainties and are consistent with both  theoretical predictions.
\begin{table}[H]
\begin{center}
\begin{tabular}{c||c|c|c||c|c|c}
\mbox{} & $|A_\perp|^2_{\rm exp}$ & $|A_\perp|^2_{FA}$ \cite{Horgan:2013hoa} & $|A_\perp|^2_{FA}$ \cite{Hambrock:2013zya} & $|A_\parallel|^2_{\rm exp}$& $|A_\parallel|^2_{FA}$ \cite{Horgan:2013hoa} & $|A_\parallel|^2_{FA}$ \cite{Hambrock:2013zya} \\  \hline
$J/\Psi K^{*}$ & $0.213 \pm 0.007$ &  $0.22 \pm 0.07$ & $0.21 \pm 0.05$ & $0.219 \pm 0.008$ & $0.39\pm 0.10$ &  $0.14 \pm 0.01$\\
$\psi(2S) K^{*}$  & $0.30 \pm 0.06$ &  $0.21\pm 0.04$ & $0.29 \pm 0.07$  & $0.22\pm 0.06$ & $ 0.48\pm 0.09$ & $0.24 \pm 0.03$  \\
\end{tabular}
 \caption{ $K^*$-polarization fractions from data with statistical and systematic uncertainties added in quadrature \cite{Aubert:2007hz,Aaij:2013cma}  and in factorization with form factors extrapolated  from the lattice  \cite{Horgan:2013hoa} and a phenomenological fit (SE2LEL)  \cite{Hambrock:2013zya}.}\label{tab:Kstpol}
 \end{center}
\end{table}  
The  factorization predictions work for the perpendicular polarization fraction, but exhibit larger spread and uncertainties in the other two, in particular for the 
$J/\psi$ final state.
The spread in theory predictions  points to the sensitivity to form factor  predictions, which  in this intermediate $q^2$ region need to be extrapolated which 
brings in additional uncertainties.

\end{appendix}

\end{document}